\documentclass[11pt,reqno]{amsart}

\usepackage{amsmath, amsfonts, amssymb, amscd}

\usepackage{amsmath, amsfonts, amssymb, amscd, enumerate}
\theoremstyle{plain}
\newtheorem{theo}{Theorem}[section]

\newtheorem{prop}[theo]{Proposition}

\theoremstyle{definition}
\newtheorem{rem}[theo]{Remark}

\newtheorem{definition}[theo]{Definition}
\newenvironment{pf}{\noindent{\it Proof. }}{$\square$\par\medskip}

\theoremstyle{plain}

\newtheorem{theorem}[theo]{Theorem}

\theoremstyle{definition}

\renewcommand{\=}{:=}

\newcommand{\beq}{\begin{equation}}
\newcommand{\eeq}{\end{equation}}
\renewcommand{\a}{\alpha}
\renewcommand{\b}{\beta}

\newcommand{\e}{\epsilon}
\newcommand{\ve}{\varepsilon}

\newcommand{\g}{\gamma}
\newcommand{\h}{\eta}

\renewcommand{\k}{\kappa}
\renewcommand{\l}{\lambda}
\renewcommand{\o}{\omega}
\newcommand{\q}{\vartheta}
\renewcommand{\r}{\rho}
\newcommand{\s}{\sigma}

\renewcommand{\O}{\Omega}

\newcommand{\bR}{\mathbb{R}}

\renewcommand{\gg}{\mathfrak{g}}

\renewcommand\sp{\mathfrak{sp}}

\newcommand{\cC}{\mathcal{C}}
\newcommand{\cD}{\mathcal{D}}

\newcommand{\cI}{\mathcal{I}}

\newcommand{\cL}{\mathcal{L}}

\newcommand{\cS}{\mathcal{S}}

\newcommand{\cU}{\mathcal{U}}

\newcommand{\cW}{\mathcal{W}}

\newcommand{\p}{\partial}

\renewcommand{\square}{\kern1pt\vbox
{\hrule height 0.6pt\hbox{\vrule width 0.6pt\hskip 3pt
\vbox{\vskip 6pt}\hskip 3pt\vrule width 0.6pt}\hrule height0.6pt}\kern1pt}

\DeclareMathOperator\Id{Id}

\DeclareMathOperator{\Span}{Span}

\newcommand\Hom{\operatorname{Hom}}

\newcommand{\wt}{\widetilde}
\newcommand{\wh}{\widehat}

\newcommand{\be}{\begin{equation}}
\newcommand{\ee}{\end{equation}}

\def\<#1,#2>{\langle\,#1,\,#2\,\rangle}
\newcommand{\arr}{\begin{array}{rlll}}
\newcommand{\ea}{\end{array}}
\newcommand{\bea}{\begin{eqnarray}}
\newcommand{\eea}{\end{eqnarray}}
\newcommand{\bean}{\begin{eqnarray*}}
\newcommand{\eean}{\end{eqnarray*}}
\newcommand{\vv}{\operatorname{v}}

\def\sideremark#1{\ifvmode\leavevmode\fi\vadjust{
\vbox to0pt{\hbox to 0pt{\hskip\hsize\hskip1em
\vbox{\hsize3cm\tiny\raggedright\pretolerance10000
\noindent #1\hfill}\hss}\vbox to8pt{\vfil}\vss}}}

\newcounter{ssig}
\newcounter{ttig}
\newcommand{	\ad}{\operatorname{ad}}
\renewcommand{\vv}{\text{\bf v}}
\newcommand{\Symm}{\Sigma}
\newcommand{\Triv}{\mathfrak{Triv}}
\newcommand{\First}{\mathfrak{ConstMot}}
\newcommand{\FirstEl}{\mathfrak{I}^{\text{elem}}}
\newcommand{\Const}{\mathfrak{Const}}
\newcommand{\Null}{\mathfrak{Null}}

\title[Lie algebras of conservation laws]
{Lie algebras
of conservation laws of  variational  ordinary differential equations}
\author[E. Fiorani and A. Spiro]
{Emanuele Fiorani  and Andrea Spiro}


\thanks{{\it Acknowledgments}. This research was partially supported by the Project MIUR ``Real and Complex Manifolds: Geometry, Topology and  Harmonic Analysis'' and by  GNSAGA of INdAM}

  \address
{\newline Emanuele Fiorani and Andrea Spiro, 
Scuola di Scienze e Tecnologie, Universit\`a di Camerino, Via Madonna delle Carceri 9,
I-62032 Camerino (Macerata),
ITALY\newline
\phantom{a}}

\email
{emanuele.fiorani@unicam.it}\par
\medskip

 \email
{andrea.spiro@unicam.it}\par
\medskip

\begin{document}
\begin{abstract} We establish a new version of the first Noether Theorem, according to which the  (equivalence classes of)  first integrals of given
Euler-Lagrange equations  in one independent variable are in  exact one-to-one correspondence
with the (equivalence classes of) vector fields satisfying    two simple geometric conditions, namely they  simultaneously preserve   the holonomy distribution of the jets space and the action  from which the Euler-Lagrange equations are derived.
\end{abstract}

\maketitle
\setcounter{section}{0}
\setcounter{subsection}{1}
\section{Introduction}
\setcounter{equation}{0}
 The first  Noether Theorem  is  surely one of the  most  celebrated  and widely studied  results  on conservation laws: see,  for instance, \cite{No, Ol, Ko, Ol2} and references therein.  As far as we know, the strongest and most general version  of this theorem has been  given  by Olver in  \cite{Ol, Ol1}. There,   in a very clear and precise way,   Olver    shows that there   exists   an exact   one-to-one   correspondence  between the family  of (equivalence classes of) conservation laws
 for  given  Euler-Lagrange equations on sections of a bundle $\pi: E \to M$, and
  the collection  of  (equivalence classes of) some special vector fields,  called {\it generalized infinitesimal symmetries},    defined on the     bundle $\pi^\infty:   J^\infty(E) \longrightarrow M$ of the infinite jets of  sections of $E$.
 \par
 \smallskip
We now recall   that any   jet bundle  of finite order $\pi^k:   J^k(E) \longrightarrow M$  is   completely  determined, up to local equivalences,   by   the pair $(N, \cD)$, formed by:
\begin{itemize}
\item[--] the total manifold $N:=J^k(E)$ of the jet bundle;
\item[--] a   special   distribution $\cD \subset T  N$,    called   {\it canonical differential system} or {\it holonomy distribution}  (\cite{Ya,Ya1, Sp}).
\end{itemize}
Indeed,  by a result by  Yamaguchi,
the pair $(N, \cD)$  characterizes  the  bundle  $\pi^k: J^k(E)\longrightarrow M$ in the following sense:   if   $(N', \cD')$ is another pair, formed by a manifold $N'$ of $\dim N' =  \dim N$ and  a non-integrable distribution
$ \cD' \subset TN'$ on $N'$,   satisfying    an appropriate   set  of conditions,  then there exists a local diffeomorphism between $N'$ and $N= J^k(E)$, which  maps   $\cD'$ into   $\cD$  and allows to  consider locally $N'$ as a jet bundle of order $k$ (\cite{Ya1}, Thm. 2.4').\par
\medskip
It is therefore natural to expect that Olver's correspondence between  conservation laws and  generalized infinitesimal symmetries  might  admit an equivalent formulation   in terms  of    vector fields on the jet bundle satisfying the following  simple conditions:  their  local flows   preserve  \par
\begin{itemize}
\item[a)] the holonomy distribution  $\cD$ and
\item[b)]   the  action $\cI$, from which   the   Euler-Lagrange equations are derived.
\end{itemize}
Such alternative    formulation of Noether-Olver's correspondence  is actually possible. \par
\smallskip
 In this paper,  we prove it   for   Euler-Lagrange equations in one independent variable. The proof  for the general case  of  equations in several  independent variables will appear  in a forthcoming paper (\cite{FGS}; see also \cite{Ge}). \par

 \medskip

 Let us call {\it infinitesimal symmetries for the action $\cI$}, or shortly {\it $\cI$-symmetries}, the vector fields of a jet bundle $J^k(E)$,  satisfying conditions (a) and (b). Our result indicates  that
the correspondence between $\cI$-symmetries and conserved quantities (better to say, {\it constants of motion}, depending on derivatives up to a fixed finite order, possibly higher than the order of the system),   is  an almost perfect analogue of the well-known    bijection  between first integrals of  a time-independent Hamiltonian system and   the  Hamiltonian     vector fields  that preserve the Hamiltonian function $H$  (see  e.g. \cite{MR}, \S 5.5). \par
\smallskip
However,   this analogy breaks down in the following  crucial  aspect.   First of all, we stress the fact that  the above correspondence is established  for {\it any} system of Euler-Lagrange equations,  derived by some variational principle.  In particular,   it   equally  applies  to  both  Lagrangian  and Hamiltonian settings. Hence, one can    explicitly apply our construction to determine   the $\cI$-symmetries associated with  the first integrals of a time-independent  Hamiltonian system that depend just on phase space  coordinates (to distinguish them from all other constants of motion, we call them {\it first integrals of elementary type}). Comparing them   with the Hamiltonian vector fields associated with such first integrals,  one can realize the following   somehow unexpected  fact: the $\cI$-symmetries and Hamiltonian vector fields  are  different objects, even though there exists a very  natural  bijection  between them. There is  however a very simple reason   behind such a  difference:    an  Hamiltonian vector field  corresponds   to a    first integral of elementary type (determined up to a constant) by means of  a  contraction with the canonical symplectic 2-form of the phase space;   an $\cI$-symmetry corresponds to a first integral of the same kind by means of  a contraction with the Poincar\'e-Cartan 1-form  of the Hamiltonian system (see \S \ref{section43} for details).   \par
\smallskip
On the basis of this fact,  our alternative presentation  of the  correspondence between  conservation laws  and  $\cI$-symmetries  can be considered as  the natural generalization of  the correspondence   between  first integrals of elementary type  and infinitesimal symmetries of  a Poincar\'e-Cartan 1-form, and not of the canonical symplectic form.  \par
In addition,  the explicit details of our proof show   the following      facts:
\begin{itemize}
\item[1)] For any  $k \geq 0$ and for any action $\cI$ on curves $\g: I \subset \bR \to E$, determined by a Lagrangian which  depends on the $k'$-th order jets of such curves with $k' \leq \left[\frac{k}{2}\right]-1$, there exists  at least one  1-form,  which is a    natural  analogue of the  Poincar\'e-Cartan 1-forms of Hamiltonian systems (we  call it   1-form  {\it of Poincar\'e-Cartan type}).
\item[2)] For a generic action $\cI$, there exist several (not just one!)  associated $1$-forms of Poincar\'e-Cartan type and  the explicit correspondence between $\cI$-symmetries and constants of motion  does depend on the choice of one such $1$-form.  It is   only   the  associated    map between  equivalence classes of $\cI$-symmetries   and of conservation laws, which  is independent of this choice.
\item[3)] For any fixed  $u \in J^k(E)$, the collection $\gg^{\cI}$ of germs at $u$ of $\cI$-symmetries has a natural structure of an infinite-dimensional Lie algebra, determined by  the usual Lie brackets between vector fields. However,  in general, the Lie algebra structure of $\gg^\cI$  {\it does not} induce a natural  Lie algebra structure on the  space $\First$ of germs of  (locally defined) constants of motion.  One can impose a corresponding natural Lie algebra structure on certain subspaces of $\First$ only  if  special restrictions are considered, as for instance if one consider  only  Hamiltonian systems  and    first integrals of elementary type.  Nonetheless,   there always  exists  a natural linear representation  of $\gg^\cI$ on $\First$, which makes $\First$  a $\gg$-module (see \S \ref{section33} below,  for  details).
\end{itemize}
\par
\smallskip
We observe that our construction  of the  correspondence between $\cI$-symmetries and conservation laws
makes use  only of classical operators of Differential Geometry,  like e.g.    exterior differentials, Lie derivatives  etc., and it has been   designed to admit simple and    direct   generalizations to  Euler-Lagrange equations on  supermanifolds.   We  plan  to undertake this task in a future paper.
\par
\medskip

As a conclusive remark, we remind that Noether theorems have a long story, clearly exposed in Kosmann-Schwarzbach's book \cite{Ko} and summarized also in Olver's review \cite{Ol2}. In \cite{Ko}, p. 143-144, the author stresses the clarity and completeness of Olver's presentation in \cite{Ol} and suggests further investigations towards other kinds of geometrical approaches to Noether theorems (see, for instance \cite{Ko0}). In our opinion, the results of this paper may be considered as a contribution in this direction.
\par
\medskip

The paper is structured as follows.
In \S 2, we introduce  the definition of  the holonomic distribution of a jet bundle $J^k(E)$, associated with  a bundle $\pi: E \to \bR$ with 1-dimensional basis,  and  of {\it variational equivalences} between  $p$-forms on $J^k(E)$.  The interest for such  equivalence relations  is motivated by the following  facts:
\begin{itemize}
\item[i)] a Hamiltonian or Lagrangian   action  $\cI$ on curves $\g: I \subset \bR \to E$, depending on the $k$-order jets of  these curves,   can be always defined as the integral of a $1$-form of $J^k(E)$ along the traces in  $  J^k(E)$ of the  curves of jets  $t \mapsto j^k(\g)|_t$;
\item[ii)] two $1$-forms   on $J^k(E)$ determine the same action $\cI$ if and only if they are variationally equivalent;
\item[iii)]  the Euler-Lagrange equations, which characterize the stationary curves  for   $\cI$,  are given by the  components of a special 2-form, which is variationally equivalent to the  exterior differentials of  the (variationally equivalent) 1-forms that  determine  $\cI$.
\end{itemize}
In \S 3, we introduce the notion of   infinitesimal symmetries of an action $\cI$ and prove the advertised  correspondence between  (equivalence classes of) such infinitesimal symmetries and (equivalence classes of) constants of motion for  the   Euler-Lagrange equations of  $\cI$.
In \S 4, we  determine the infinitesimal symmetries of the action, associated with   a (time-independent)  Hamiltonian system,  and compare them
with the Hamiltonian vector fields, associated with  first integrals of elementary type. Finally,  using Darboux Theorem, we  get our final result,   Theorem \ref{theorem44}, which  generalizes a previous   theorem by Mukunda (\cite{Mu}).
\par
\medskip

\noindent
{\it Acknowledgements.} We are grateful to Franco Cardin and Wlodzimierz Tulczyjew for very useful discussions on many aspects of this paper.

  \bigskip
\section{Geometrization of Euler-Lagrange equations of one independent variable}
 \subsection{Notational remarks}
In this paper we are concerned with the systems of ordinary differential equations for curves
 $\g: I\subset \bR \longrightarrow M$ on an $n$-dimensional manifold $M$, which are Euler-Lagrange equations
 determined by some variational principle. \par
 \smallskip
Main examples of such equations are given
by the differential  systems occurring in Lagrangian and Hamiltonian mechanics. In these cases, the manifold  $M$ plays the role of   the   {\it configuration space} or {\it phase space} of the considered physical system. The parameter $t \in I \subset \bR$ of the curve  has to be considered as the  {\it time coordinate}. \par
 \smallskip
In our discussion, the 1-dimensional manifold $\bR$ is  constantly considered  with  a fixed orientation, namely the  one determined by the trivial   coordinate system $\Id_\bR =  (t): \bR$ $ \longrightarrow \bR$. The globally defined  1-form $dt$  is referred to as {\it  standard volume form} of $\bR$. \par
\smallskip
It is immediate to realise that any (smooth) parameterized curve  $\g: I\subset \bR \longrightarrow M$ is  uniquely associated with the corresponding   (local) section of the trivial bundle
$\pi: E = \bR \times M  \longrightarrow  \bR$
$$\wh \g_t \= (t,\g_t)\ .$$
So, with no loss of generality, in place of   parameterized curves in $M$,  all results of this paper   are   expressed in terms of  local (smooth) sections of the trivial bundle $(E = M \times \bR, \bR, \pi)$.   \par
\bigskip
Consider an integer $k \geq 1$. Given a local section $\g: I \to E = \bR \times M $, we  use the notation $j^k_{t}(\g)$ for the $k$-th order jet of $\g$ at $t \in I$. The space of  $k$-jets of local sections of  the  bundle $(E , \bR, \pi)$  is denoted by $J^k(E)$.  \par
\bigskip
For any $1 \leq \ell \leq k$, we indicate by  $\pi^k_\ell$
the natural projection
$$\pi^k_\ell: J^k(E) \longrightarrow J^\ell(E)\ ,\qquad \pi^k_\ell(j^k_{t}(\g)) \= j^\ell_{t}(\g)\ .$$
We  also consider the natural projections $\pi^k_0:J^k(E) \longrightarrow E$ and $\pi^k_{-1}: J^k(E) \longrightarrow \bR$, defined by
$$\pi^k_0(j^k_{t}(\g)) \= \g_{t}\ ,\qquad \pi^k_{-1}(j^k_{t}(\g)) \= t\ .$$
\par
\medskip
Given a  section $\g : I \subset \bR \longrightarrow E$,  we call  {\it  lift of $\g$ to the $k$-th order} the associated curve of jets
$$\g^{(k)}: I \subset \bR \longrightarrow J^k(E)\ ,\qquad \g^{(k)}(t) \= j^k_t(\g)\ .$$
\par
\medskip
For a given  system of coordinates $\xi = (y^i): \cU \subset M \longrightarrow \bR^n$ on $\cU \subset M$,  the coordinates on $E$, defined by
$$\wh \xi: I \times \cU  \subset E \longrightarrow  \bR^{n+1}\ ,\qquad \wh \xi(t, x) = (t, y^1(x), \ldots, y^n(x))\ ,$$
are called   {\it associated with $\xi = (y^i)$}.
In general,  any  set of  coordinates on $E$ of this form is called   {\it set of adapted coordinates}.\par
\smallskip
Given  a set of adapted coordinates $\wh \xi = (t, y^i)$ on $I \times \cU \subset E$, we may consider the naturally associated coordinates
$$\wh \xi^{(k)} = (t, y^i, y^i_{(1)}, \ldots,  y^i_{(k)})  : (\pi^k_0)^{-1}(\cU \times I) \subset J^k(E) \longrightarrow \bR^{n(k +1) +1}\ ,$$
which sends a given  $k$-th order jet $u = j^k_{t}(\g)$ into the $N$-tuple, with $N = n(k +1)+1$, defined by
$$\left(t, y^i, y^i_{(1)},\ldots,  y^i_{(k)}\right)(u) \= \left(t,  \g^i_{t}, \left.\frac{d\g^i}{ds} \right|_{s = t}, \ldots, \left.\frac{d^k\g^i}{ds^k} \right|_{s = t}\right)\ .$$
We call such coordinates a  {\it set of adapted coordinates of $J^k(E)$}.\par
\bigskip
\subsection{Holonomic distributions and variational classes}
\begin{definition} The {\it holonomic distribution} of $J^k(E)$ is the distribution $\cD \subset T J^k(E)$, which is defined at any $u  \in J^k(E)$ by
$$\cD_u = \Span\left\{\ v \in T_u J^k E\ :\ v = \left.\frac{d \g^{(k)}}{ds}\right|_{s = t}\ \text{for some}\ \phantom{aaaaaaaaaaaaaaa}\right.$$
$$\left. \phantom{aaaaaaa \left.\frac{d \g'{}^{(k)}}{dt}\right|_{t = t_o} aaaaaaaaaaa} \ \g: I \longrightarrow E\ \text{such that}\ \ j^k_{t}(\g) = u\ \right\}\ .$$
The vectors  in $\cD$ and the vector fields with  values in $\cD$ are called {\it holonomic}.
\end{definition}
\par
\bigskip
Consider a system of adapted coordinates $(t, y^i, y^i_{(1)}, \ldots, y^i_{(k)})$.  If $\g$ is a section such that    $u = j^k_{t}(\g)$, the components of  $v = \left.\frac{d \g^{(k)}}{ds}\right|_{s= t}$ along the direction of   $\left.\frac{\partial}{\partial t}\right|_u$ and $\left.\frac{\partial}{\partial y^i_{(a)}}\right|_u$,   $1 \leq a \leq k-1$, are completely fixed by
the coordinates $(y^i_{(1)}, \ldots, y^i_{(k)})$. The other components  are not  determined by the coordinates of  $u$ and may vary arbitrarily. From these  observation,  a   basis for  the subspace   $\cD_u \subset T_u J^k(E)$  is  given by the vectors
\beq\left.\frac{d}{dt}\right|_{u = (t, y^i_{(a)})} \!\!\!\!\!\!\!\!\=\left.\frac{\partial}{\partial t}\right|_u + \sum_{a=0}^{k-1} y^j_{(a+1)} \left.\frac{\partial}{\partial y^j_{(a)}}\right|_u\qquad \text{and}\qquad  \left.\frac{\partial}{\partial y^i_{(k)}}\right|_u\ ,\qquad 1 \leq i \leq n\ , \eeq
and the vector fields $\frac{d}{dt}$,  $\frac{\partial}{\partial y^i_{(k)}}$ is a collection of  local generators for $\cD$.\par
\bigskip
\begin{definition} A locally defined $p$-form $\l$ of $J^k(E)$, $p \geq 1$,  is called  {\it holonomic} if $\imath_X \l = 0$
for any holonomic vector field $X$.  A local $0$-form  (i.e., a  $\cC^\infty$ function on an open set) is called {\it holonomic} if it vanishes identically.\par
If $\a$, $\a'$ are $p$-forms  on the same open subset $\cU \subset J^k(E)$,
they are called {\it variationally equivalent} if
$$ \a - \a' = \l + d \mu $$
for some holonomic $p$-form  $\l$ and some holonomic $(p-1)$-form  $\mu$.
\end{definition}
\par
\medskip
By previous remarks,  given a set of adapted coordinates $\wh \xi^{(k)} = (t, y^i_{(a)})$, the holonomic 1-forms are exactly those that are  linear combination of the 1-forms   (here, $y^i_{(0)}\= y^i$)
\beq \label{holonomicforms} \o^i_{(a)} \=  dy^i_{(a)} - y^i_{(a+1)} dt\ , \qquad a  = 0, \ldots, k-1,\eeq
at all points.\par
\smallskip
Note also that  if $\mu$ is holonomic, its differential $d \mu$ might be non-holonomic. For instance,  the 1-forms
$\o^i_{(a)}$, $a \leq k-2$,  are holonomic, but their differentials are of the form $d\o^i_{(a)} = dy^i_{(a+1)} \wedge dt = \o^i_{(a+1)} \wedge dt$ and are not holonomic. Indeed,
$$\imath_{\frac{d}{dt}} d \o^i_{(a)}= - \o^i_{(a+1)}  \neq 0\ .$$
\par
\medskip
The relation of variational equivalence is an equivalence relation  between   $p$-forms  defined on the same  open subset $ \cU \subset J^k(E)$.  If  $\a$ is  a $p$-form on $\cU$, we call {\it variational class of $\a$} the collection $[\a]$ of all $p$-forms that are variationally  equivalent to $\a$.\par
\medskip
The main motivation  for considering the notion of variational classes  is discussed  in the next section.\par
\bigskip
\subsection{Actions, Lagrangians and Poincar\'e-Cartan forms}
\label{sec2.3}
As mentioned in the Introduction, we are  concerned with  conservation laws for Hamiltonian systems  as well as
for any set of Euler-Lagrangian equations on sections $\g: I \longrightarrow E = M \times \bR$,  originating from functionals   of the form
\beq\label{action1} I_L(\g) = \int_{a}^b L(\g^{(k)}) dt\ . \eeq
Here  $L: J^k(E) \longrightarrow \bR$ denotes a $k$-th order Lagrangian, that is a  $\cC^\infty$ real  function on $J^k(E)$. \par
\medskip
For such purposes, it is very convenient to   consider the following notion.
\par
\begin{definition}  Let  $[\a]$ be the variational class of a 1-form $\a$ on $J^k(E)$. We call {\it action associated with  $[\a]$}
the functional
\beq\label{action2}  \cI_{[\a]}:\left\{\mbox{local sections}\  \g:I\subset \bR\longrightarrow E\right\}\longrightarrow \bR\ ,$$
$$ \cI_{[\a]}(\g)\=\int_{\g^{(k)}(I)}\a\eeq
(in this formula,  $\int_{j^k (\g)(I)}\a$ indicates    the integral  of   $\a$ along the  1-dimensional submanifold  $\g^{(k)}(I) \subset J^k(E)$).
\end{definition}
\par
\bigskip
Here is a   sequence of remarks that motivate this  definition.\par
\  \\
(1) The functional $\cI_{[\a]}$ is well defined. Namely,   if $\a$, $\a'$ are such that $[\a'] = [\a]$ then $\a' = \a + \l$ for some holonomic  $\l$  and
$$\int_{\g^{(k)}(I)}\a' = \int_{\g^{(k)}(I)}\a + \int_{\g^{(k)}(I)}\l = \int_{\g^{(k)}(I)}\a\ ,$$
since $\l$ is $0$ on all vectors    tangent  to $\g^{(k)}(I)$, because they  are holonomic.\\
\ \\
(2) For any Lagrangian $L: J^k(E) \longrightarrow \bR$,  the 1-form $\a_L \= L \cdot (\pi^k_{-1})^*dt$ is such that,  for any  section $\g$,
$$\cI_{[\a_L]} (\g)= \int_{\g^{(k)}(I)}\a_L = \int_{a}^b L(\g^{(k)}) dt = I_L(\g) \ . $$
This shows   that  \eqref{action1}  coincides  with  the action associated with  $[\a_L]$. \\
\ \\
(3) Let $\a$ be a 1-form on $J^{k}(E)$ and denote by $\wt \a = (\pi^{k+1}_k)^*(\a)$ the pull-back of $\a$ on the jet space $J^{k+1}(E)$.
Let also  $\cW \subset J^{k+1}(E)$ be  an open subset admitting a set of adapted coordinates $\wh \xi^{(k)} = (t, y^i, y^i_{(a)})$.  The collection of $1$-forms $(dt, \o^i_{(a)}, dy^j_{(k+1)})$
is  a coframe field on $\cW$ and any $1$-form is a linear combination of such $1$-forms at any point.  Since
$$\imath_{\frac{\partial}{\partial y^i_{(k)}}}\wt \a =  \a\left(\pi^{k+1}_k{}_*\left(\frac{\partial}{\partial y^i_{(k)}}\right) \right) = 0\qquad \text{for all} \ 1 \leq i \leq n\ ,$$
it follows  that   $\wt \a|_{\cW}$ has trivial  components along  the 1-forms like $dy^j_{(k+1)}$.  It is therefore  of the form
 $$\wt \a|_{\cW} = L dt +  \sum_{a = 0}^k \a_{i(a)} \o^i_{(a)}\ ,$$
for some smooth real functions $L$, $\a_{i(a)}$ on $\cW$. Since $\sum_{a = 0}^k \a_{i(a)} \o^i_{(a)}$ is holonomic and $dt$ coincides with   the pull-back of the standard volume form $dt$ of $\bR$, we conclude  that
$[\wt \a|_{\cW} ] = [ \a_L]$
and the values  of the functional $\cI_{[\a]}$ on  sections  of $\cW$ are given by
$$\cI_{[\a]}(\g) = \int_{\g^{(k+1)}(I)}\wt \a = \int_a^b L(\g^{(k+1)}) dt = I_L(\g)\ .$$
This means that, locally, {\it   $\cI_{[\a]}$  can be always identified with a functional  of the form  $I_L$,  given by an appropriate $(k+1)$-th order Lagrangian $L$}. \\
\ \\
(4) Let $M = T^* \bR^N$ be the phase space of a classical mechanical system and $H: T^* \bR^N \longrightarrow \bR$ the Hamiltonian, which determines the
dynamics of the system. As it is well known, the Hamilton equations $\dot q^i =  \frac{\partial H}{\partial p_i}$, $\dot p_j = -  \frac{\partial H}{\partial q^j}$ are the Euler-Lagrange equations that arise from a  variational principle on the action
\beq \label{action3}ÊI(\g) = \int_{\g(I)} p_i dq^i - H dt\eeq
on  sections of $\pi: (T^* \bR^N) \times \bR \longrightarrow \bR$. The 1-form $\a_H  = p_i dq^i - H dt$ is  usually called  {\it Poincar\'e-Cartan form}.  Note that  \eqref{action3} coincides with   the action $\cI_{[\a_H ] }$  associated with the   variational class $[\a_H]$.\par
\bigskip
By these observations, it is clear that  the  actions determined by variational classes  constitute  a set that naturally  includes and extends the class of all   actions in    Lagrangian and Hamiltonian mechanics.   With the purpose of  dealing with   both kinds of  such actions on  the same  footing, from now on our discussion  is done in the general terms  of  variational classes and associated actions.
\par
\medskip
We conclude with a  very convenient  definition. \par
\begin{definition} A $p$-form $\wt \b$ on (an open subset of) $J^k(E)$  is said  {\it of order r} for some $0 \leq r \leq k$   if there exists a $p$-form $\b$ on (an open subset of) $J^r(E)$ such that $\wt \b = (\pi^k_{r})^* \b$.
\end{definition}
Using this definition,  by  a  pull-back,   {\it a $p$-form $\a$  on  $J^r(E)$ can be  considered as {\rm $p$-form  of order $r$}  on a jet space $J^k(E)$  for any   $k\geq r$}.  \par
\bigskip
\subsection{Variational Principles and Euler-Lagrange equations}
\label{section2.4}
We now want  to  introduce  a definition of  variational principles for actions given by   variational classes, which  directly   implies the
usual   Euler-Lagrange equations in  Lagrange or Hamiltonian settings.  For  this, we first  need to consider the following
generalized  definition  of {\it variation with fixed boundary}. \par
\medskip
 Let $\g: I \longrightarrow E = M \times \bR$ be a local section and $[a, b] \subset I$ a closed subinterval of its domain $I$.
We call  {\it smooth variation of $\g$ with fixed  boundary  up to order $k$ } any  smooth  map $F: [a,b] \times [-\ve, \ve] \subset \bR^2\longrightarrow M$,   such that:
\begin{itemize}
\item[a)]  all maps $\g^{(s)} \= F(\cdot, s): [a, b] \to E$,  $s \in [-\ve, \ve]$,  admit $\cC^\infty$ extensions  $\g^{(s)}: I^\ve \longrightarrow E$ on intervals $I^\ve \supset [a,b]$, which are  sections of $E$;
\item[b)]$\g^{(0)} = \g$;
\item[c)]  the $k$-th order jet curves $(\g^{(s)} )^{(k)} : [a,b] \longrightarrow J^k(E)$ are such that
$$(\g^{(s)} )^{(k)}(a) = \g^{(k)}(a)\ ,\qquad (\g^{(s)} )^{(k)}(b) = \g^{(k)}(b)$$
for all  $s \in [-\ve, \ve]$.
\end{itemize}
\medskip

\begin{definition}\label{variational}
Let $\g:I\subset \bR\longrightarrow E = M \times \bR$ be a   section and  $\cI_{[\a]}$ the action determined by a 1-form $\a$ of order $r$ in $J^k(E)$. We say that  $\g$ {\it satisfies the variational principle  determined by   $\cI_{[\a]}$} if
\beq \label{stationarity} \left.\frac{d \cI_{[\a]}(\g^{(s)})}{ds}\right|_{s=0}=0\ ,\qquad \g^{(s)} \= F(\cdot, s)\ ,\eeq
for all  smooth variations $F$   with fixed boundaries  up to order $r$  of the restrictions $\g|_{[a,b]}$ on  all closed subintervals $[a,b] \subset I$.
\end{definition}
Condition \eqref{stationarity}  clearly depends only on the first order jet in the variable $s$ of
the  variation $F$.  Indeed it is equivalent to a condition which involve some special vector fields, which  we now introduce. \par
\medskip
 Let $\g: I \longrightarrow E = M \times \bR$ be a section and
 $$W: \g^{(k)}([a,b])\longrightarrow T J^k(E)|_{\g^{(k)}([a,b])}$$
 a vector field,  which is defined  only at the points  of $\g^{(k)}([a,b])$,   $[a,b] \subset I$.   We say that  $W$ is  a {\it $k$-th order variational field}  if there exists a smooth variation $F: [a,b]\times [-\ve,\ve]\longrightarrow E$ of $\g$   with fixed boundary  up to order $k$,   such that
 \beq \label{variationalfield} W= F^{(k)}_*\left({\left.\frac{\p}{\p s}\right|}_{(x,0)}\right)\ ,\eeq
where $F^{(k)}$  is the map
 $$F^{(k)}:[a,b]\times [-\e,\e]\longrightarrow J^k(E)\quad ,\qquad  F^{(k)}(t,s)=(\g^{(s)})^{(k)}(t) = j^k_t(F(\cdot,s))\ .$$
\par
\medskip
\begin{prop}\label{TeoGusteaux}
A section $\g:I\longrightarrow E$ satisfies the variational principle determined by $\cI_{[\a]}$ if and only if for any closed subinterval $[a,b] \subset I$ and any $k$-th order variational field $W$ at the points of $\g^{(k)}([a,b])$,
\beq \label{infinitesimalstationarity} \int_{\g^{(k)}([a,b])}\imath_W d \a = 0\ .\eeq
\end{prop}
\begin{pf}  Let $J = [a,b] \subset I$ and denote by $F$  a smooth variation  with fixed boundaries  up to order $r$  of   $\g|_J$. We also indicate by $W$ the variational field along $\g|_{[a,b]}$, which is determined by $F$   by means of  \eqref{variationalfield}. By  Stokes Theorem and the conditions satisfied by  $F$ at the points $(a,s)$ and $(b, s)$,
$$ \left.\frac{d \cI_{[\a]}(\g^{(s)})}{ds}\right|_{s=0} = \lim_{h\to0} \frac{1}{h} \left( \int_{(\g^{(h)})^{(k)}(J)} \a  -  \int_{(\g^{(0)})^{(k)}(J)}\a\right) = $$
$$ =  \lim_{h\to0} \frac{1}{h} \left( \int_{F(J \times [0,h])}d\a\right) = \int_{\g^{(k)}([a,b])}\imath_W d \a\ .$$
From this, the claim follows.
\end{pf}
\par
\bigskip
At a first glance, condition \eqref{infinitesimalstationarity}  looks difficult to be handled, because it involves  the notion of
variational vector fields, which are objects that    might be hard to   characterize in terms of  explicit differential equations. \par
\smallskip
On the other hand,   we  observe that \eqref{infinitesimalstationarity}  is satisfied if and only if  $ \int_{\g^{(k)}([a,b])}\imath_W \b = 0$ for any  $\b  \in [d\a]$. Indeed, if $\b= d\a + \l + d\mu$,  for some holonomic $\l$ and $\mu$,
\beq \label{infinitesimalstationarity1}  \int_{\g^{(k)}([a,b])}\imath_W d \a \overset{\l\ \text{is holonomic}}= \int_{\g^{(k)}([a,b])}\imath_W \b -
\int_{\g^{(k)}([a,b])}\imath_W d\mu  \overset{\text{Stokes Thm.}}= $$
$$ =  \int_{\g^{(k)}([a,b])}\imath_W \b -  \mu(W)|_{\g^{(k)}(b)} +\mu(W)|_{\g^{(k)}(a)} \overset{\text{fixed boundary}}=   \int_{\g^{(k)}([a,b])}\imath_W \b\ .\eeq
By this fact, it turns out that it is very convenient to consider the following kind of 2-forms, which, as we will shortly see,  lead naturally
to the  Euler-Lagrange equations of the considered variational principle. \par

\begin{definition} \label{defsource} A $2$-form $\s$ on  $J^k(E)$, $k \geq 1$  is called  {\it source form}  if
\beq \label{sourceform} \imath_V \s = 0\eeq
for any vector field $V$ such that $(\pi^k_0)_*(V) = 0$.
\end{definition}
If $\xi^{(k)} = (t, y^i, y^i_{(1)}, \ldots y^i_{(k)})$ is  a system of adapted coordinates on an open subset  $\cU \subset J^k(E)$, condition \eqref{sourceform} is equivalent  to the equations
\beq \label{2.11}\s\left({\frac{\partial}{\partial y^i_{(a)}}}, \cdot\right) = 0\ ,\qquad 1 \leq a \leq k\ ,\eeq
to be satisfied at all points of $\cU$.  It follows that  $\s$ satisfies \eqref{sourceform}  if and only if it is of the form
$$\s  = \wt \s_j dy^j \wedge dt + \sum_{k< \ell} \wt \s_{k\ell} dy^k \wedge dy^\ell= $$
$$ = \left(\wt \s_j +  \sum_{k < \ell}(\wt \s_{k\ell} y^\ell_{(1)}- \wt \s_{k\ell} y^k_{(1)}) \right)\o^i_{(0)} \wedge dt + \sum_{k < \ell} \wt \s_{k\ell}\o^k_{(0)} \wedge \o^\ell_{(0)}   $$
for some smooth functions $\wt \s_j$ and $\wt \s_{k\ell}$. \par
\medskip
Coming back to \eqref{infinitesimalstationarity} and \eqref{infinitesimalstationarity1}, by  \cite{Sp} Prop. A.2, {\it if  $\a$ is a 1-form, which is locally variationally equivalent to a 1-form $L dt$ of order $r$,  and it is considered  (through a pull-pack) as a $1$-form on $J^k(E)$, with $k \geq 2r$,
   the class  $[d\a]$ on $J^k(E)$ contains  exactly one  source form $\s \in [d \a]$, which has  the expression
 \beq \label{2.12} \s= \s_i \o^i_{(0)} \wedge dt\eeq
in any set of adapted coordinates.}\par
  \medskip
 For reader's convenience, we show the existence of a source form  as above  in the simple case,  in which $\a $ is  defined on an open set $\cU \subset J^k(E)$, $k \geq 2$, endowed with adapted coordinates $\xi^{(k)} = (t, y^i_{(a)})$,  and it is already of the form $\a = L dt$ for some Lagrangian $L$ of order $1$.  In this case,
 $$d \a =  \frac{\partial L}{\partial y^i} d y^i \wedge dt  + \frac{\partial L}{\partial y^i_{(1)}} d y^i_{(1)} \wedge dt = \frac{\partial L}{\partial y^i} \o^i_{(0)}\wedge dt  + \frac{\partial L}{\partial y^i_{(1)}} d \o^i_{(0)} = $$
 $$ = \frac{\partial L}{\partial y^i} \o^i_{(0)} \wedge dt  + d \left(\frac{\partial L}{\partial y^i_{(1)}} \o^i_{(0)}\right) -d \left(\frac{\partial L}{\partial y^i_{(1)}}\right) \wedge \o^i_{(0)} = $$
 $$ = \left(\frac{\partial L}{\partial y^i} - \frac{d}{dt} \left(\frac{\partial L}{\partial y^i_{(1)}}\right)\right)\o^i_{(0)} \wedge dt + d \left(\frac{\partial L}{\partial y^i_{(1)}} \o^i_{(0)}\right) - $$
 $$ -  \left(\frac{\partial^2 L}{\partial y^j \partial y^i_{(1)}}\right)\o^j_{(0)} \wedge \o^i_{(0)} -  \left(\frac{\partial^2 L}{\partial y^j_{(1)}\partial y^i_{(1)}}\right)\o^j_{(1)} \wedge \o^i_{(0)}\ . $$
Since $\mu = \frac{\partial L}{\partial y^i_{(1)}} \o^i_{(0)}$ and $\l_{(a)} \= -  \left(\frac{\partial^2 L}{\partial y^j_{(a)}y^i_{(1)}}\right)\o^j_{(a)} \wedge \o^i_{(0)}$, $a =0,1$,  are holonomic,  we see that the variational class $[d \a]$ contains the  source form
\beq \label{cappa} \s = \s_i \o^i_{(0)} \wedge dt \qquad \text{with}\qquad \s_i =  \frac{\partial L}{\partial y^i} - \frac{d}{dt} \left(\frac{\partial L}{\partial y^i_{(1)}}\right)\ .\eeq
\par
\medskip
We are now able to prove that  that the sections  which satisfy a variational principle,  are  exactly the solutions  of  an appropriate
system of Euler-Lagrange equations, as  expected.
\begin{theorem}\label{cor2.3.4}  Assume that  $\a$ is a 1-form on  a jet space $J^k(E)$, which is (locally)  variationally equivalent to some  form of order $r$ of the kind $L dt$  for some   $r \leq \frac{k}{2}$. Let also $\s$ be a  source form in  $[d\a]$.  A section  $\g:I\longrightarrow E$
satisfies the variational principle of $\cI_{[\a]}$ if and only if
\beq \label{Euler-Lagrange} \imath_{\dot \g^{(k)}_t} \s= 0\qquad \text{for any}\ \ t \in I\ .\eeq
\end{theorem}
\begin{pf} First of all, we observe that if $\s$ and $\s'$ are  source forms in the same variational class $[d \a]$, i.e., such that $\s - \s' = \l + d \mu$ for some holonomic  $\l$ and  $\mu$, then  $d \mu $ is holonomic and the whole difference $\s - \s' $ is holonomic. In fact,  if   $d \mu \neq 0$ and not holonomic,  in some set of adapted coordinates  $d \mu$ is necessarily  of the form
$$d \mu = \sum_{1 \leq a}^k \mu^a_i d y^i_{(a)} \wedge dt$$
for some non-trivial functions $\mu^a_i$.
But this would contradict the fact that $\s$ and $\s'$ are both source forms, hence both satisfying \eqref{2.11}. Due to this and the fact that,  for any section $\g$,   the tangent vectors  $\dot \g^{(k)}_t$ are in $\cD$,
 we get  that  $\imath_{\dot \g^{(k)}_t} \s= \imath_{\dot \g^{(k)}_t} \s'$.\par
By this remark,  with no loss of generality,  from now on we may  assume that $\s$ is the unique
source form of $[d\a]$ described  in \eqref{2.12}.
By \eqref{infinitesimalstationarity1} and Proposition \ref{TeoGusteaux}, $\g$ satisfies the variational principle if and only if
\beq \label{acca} \int_{\g^{(k)}([a,b])}\imath_W \s = 0\eeq
for any closed subinterval $[a,b] \subset I$ and any $k$-th order variational field $W$. If we consider $[a,b]$ so small so that $\g^{(k)}|_{[a,b]}$ is included in the domain
of a  system of adapted coordinates $\xi^{(k)} = (t, y^i_{(a)})$, we have that $W$ and $\imath_W \s$ are of  the form
\beq \label{el}W = W^i \frac{\partial}{\partial y^i} + \sum_{a = 1}^k W^i_{(a)} \frac{\partial}{\partial y^i_{(a)}} \ ,\quad   \imath_W \s = \imath_W(\s_i \o^i_{(0)} \wedge dt) = (W^i \s_i) dt\ .\eeq
We now observe that  for any choice of functions $f^i: \g^{(k)}([a,b]) \to \bR$ that vanish identically on neighborhoods of $a$ and $b$, one can  construct   a smooth variation $F$  with  fixed boundary up to order $k$,  whose associated variational field $W$  satisfies
$$W^i|_{\g^{(k)}(t)}  = f^i|_{\g^{(k)}(t)}$$
at any $t \in [a,b]$. This fact together with \eqref{el} implies that \eqref{acca} is satisfied for all subintervals $[a,b]$ and all variational  fields $W$ if and only if the functions $\s_i|_{\g^{(k)}(t)}$ are identically vanishing. Since
$$ \imath_{\dot \g^{(k)}_t} \s= \left(\s_i|_{\g^{(k)}(t)}\right)\left(\o^i(\dot \g^{(k)}_t) dt - \o^i_{(0)}\right) = - \s_i|_{\g^{(k)}(t)} \o^i_{(0)}$$
the claim follows.
\end{pf}
\par
\medskip
By previous remarks and the proof of Theorem \ref{cor2.3.4}, using a set of adapted coordinates,
 the equation \eqref{Euler-Lagrange} is equivalent to the system
\beq \label{EL} \s_i(\g^{(k)}(t)) = 0\ ,\qquad 1 \leq i \leq n\ ,\eeq
where  the $\s_i$'s are  the components of the unique source form $\s \in [d\a]$ described in \eqref{2.12}. By \eqref{cappa}, when $\a$ is of the form  $\a = L dt$ for some Lagrangian $L$ of first order, the equations
\eqref{EL} are the  Euler-Lagrange equations
$$\left(\frac{\partial L}{\partial y^i} - \frac{d}{dt} \left(\frac{\partial L}{\partial y^i_{(1)}}\right)\right)_{\g^{(1)}(t)} = 0\ .$$
The reader can directly check that \eqref{EL} coincide with the   Eulero-Lagrange equations of a  Lagrangian $L$ also in the cases in which  $L$ is of order higher than one.\par
\bigskip
\par
\medskip

\section{A geometric proof of Noether Theorem for variational systems of o.d.e.'s}
\setcounter{equation}{0}
\subsection{Conservation laws and symmetries of variational o.d.e.'s}
Let $\a$ be a 1-form on $J^k(E)$ and $f: \cU \subset J^k(E) \longrightarrow \bR$  a smooth function, defined on an open subset of $J^k(E)$.
\begin{definition} The function $f$ is  said {\it constant of motion} for the  variational principle of   $\cI_{[\a]}$ if
for any  section $\g: I \longrightarrow E$ that satisfies  the variational principle,
$$\left.\frac{d (f\circ \g^{(k)})}{dt}\right|_t = 0\qquad \text{for any}\ t  \in I\ .$$
\end{definition}
\medskip
\par
As we will shortly see, the (first) Noether Theorem establishes a natural correspondence between  symmetries of  $\cI_{[\a]}$ and  conservation laws. Indeed, such correspondence  appears to be   a  bijection, provided that  the  objects  that are   called    {\it symmetries}  are   specified in an appropriate way.   To  this purpose,  the following definition is crucial. \par
\begin{definition}\label{defDsymmetries} Let  $X$ be a vector field   and $\a$ a 1-form on $J^k(E)$.
\begin{itemize}
\item[a)]$X$ is called {\it infinitesimal  symmetry  of $\cD$} (shortly,  {\it $\cD$-symmetry}) if, for any  holonomic vector field $Y$,  the  Lie derivative $\cL_XY$ is also a holonomic vector field.
\item[b)]$X$ is called  {\it infinitesimal  symmetry  for   $\cI_{[\a]}$}  if
 it  is   a $\cD$-symmetry
and  $\cL_X \a$ is holonomic for some (and hence for all) $\a \in [\a]$.
\end{itemize}
\end{definition}
\begin{rem} As a direct consequence of definitions,     $X$ is a $\cD$-symmetry if and only if  the local flow $\Phi^X_t$ of $X$ around  any jet $u \in J^k(E)$, maps the holonomic distribution $\cD$ into itself. This implies  the following two  crucial facts:
\begin{itemize}
\item[a)] A vector field $X$ on some open subset  $\cU \subset J^k(E)$ generates a 1-parameter family  of diffeomorphisms that  transform $k$-lifts $\g^{(k)}$ of sections into curves that are  also  lifts of sections if and only if it is a $\cD$-symmetry. \par
This is the main reason of   interest for $\cD$-symmetries. \\
\item[b)] If  $X$ is a $\cD$-symmetry and $\l$ is  a holonomic $p$-form, also   the $p$-forms $\Phi^X_t{}^*\l$, $t \in ]-\ve, \ve[ \subset \bR$,   and the Lie derivative $\cL_X \l$,  are holonomic. From this, it follows   that if $\a$ and $\a'$ are variationally equivalent (i.e. $\a - \a' = \l + d \mu$, with $\l$, $\mu$ holonomic), then
$\cL_X \a$ is holonomic if and only if $\cL_X \a'$ is holonomic. \par
This explains why the definition of infinitesimal symmetry for $\cI_{[\a]}$   depends  on  the variational class $[\a]$ and not on the choice of the 1-form $\a$ in that class.\\
 \end{itemize}
 If we denote by $\cS^{(k)}$ the class of  $k$-th order lifts of sections of $\pi: E \to \bR$ and we consider $\cI_{[\a]}$ as an operator  $\cI_{[\a]}: \cS^{(k)} \longrightarrow \bR$ with domain $\cS^{(k)}$,   (a) and (b) lead to the following interpretation of the notions     in Definition \ref{defDsymmetries}.
\begin{itemize}
\item[--] {\it The local flows of  $\cD$-symmetries  can be considered as    1-parameter  groups of local transformations of $\cS^{(k)}$;}
\item[--] {\it  The local flows of  infinitesimal symmetries of $\cI_{[\a]}$ can be considered as   1-parameter  groups of local transformations of $\cS^{(k)}$, with orbits along which the functional $\cI_{[\a]}$  is constant. }
\end{itemize}
\end{rem}
\medskip
In the next  proposition, we show that the $\cD$-symmetries and the infinitesimal symmetries for an action $\cI_{[\a]}$ coincide with the   vector fields that satisfy  an appropriate system of partial  differential equations.
\begin{prop}\label{equiv-Dsimm}
Let $X$ and $\a$ be a vector field and a $1$-form on   $J^k(E)$, respectively, and  $\wh \xi^{(k)} = (t, y^i_{(a)})$ a system of adapted coordinates on $\cU \subset J^k(E)$. Then:
\begin{itemize}
\item[1)] $X|_{\cU}$  is a $\cD$-symmetry if and only if  it satisfies the following system of p.d.e.'s
\beq \label{Dsymmetrieseq}
\left\{\begin{array}{ll}\o^i_{(a)}\left(\cL_X \frac{d}{dt}\right) = 0\ , &0 \leq a \leq k-1\ , \quad 1 \leq i \leq n\ ,\\
 \\
\o^i_{(a)}\left(\cL_X \frac{\partial}{\partial y^j_{(k)}}\right)  = 0\ , &0 \leq a \leq k-1, \quad 1 \leq i, j \leq n\ .
 \end{array}\right.
 \eeq
 \item[2)] $X|_{\cU}$ is an infinitesimal symmetry for $\cI_{[\a]}$ (considered as functional  on  sections in $\cU$) if and only if it satisfies
 \eqref{Dsymmetrieseq} and the equations
  \beq \label{symmetries}
\left\{\begin{array}{l}(\cL_X \a)\left(\frac{d}{dt}\right) = 0\ , \\
 \\
(\cL_X \a)\left( \frac{\partial}{\partial y^j_{(k)}}\right)  = 0\ .
 \end{array}\right.
 \eeq
 \end{itemize}
\end{prop}
\begin{pf} We recall that $\cD|_{\cU}$ is generated by the vector fields $\frac{d}{dt}$ and $\frac{\partial}{\partial y^j_{(k)}}$, $1 \leq j \leq n$. This implies that  a vector field takes values in $\cD|_{\cU}$ if and only if it is in the intersection of the   kernels of  the 1-forms $\o^i_{(a)}$, $0 \leq a \leq k-1$. From these two facts, it follows  that $X|_{\cD}$ is a $\cD$-symmetry if and only if (1) holds and that $\cL_X \a$ is holonomic (i.e. it vanishes on all holonomic vector fields) if and only if (2) is satisfied.
\end{pf}
We conclude with an explicit   description  of   $\cD$-symmetries in adapted coordinates.  In the next statement,  $\wh\xi^{(k)} = (t, y^i,y^i_{(a)})$ is a  fixed     system of adapted coordinates
 on an open subset  $\cU \subset J^k(E)$. Moreover, for any smooth map
$\vv = (\vv^0, \vv^1, \ldots, \vv^n): \cU \subset J^k(E) \longrightarrow \bR^{n+1}$
we adopt the notation   $X_\vv$ to indicate the   vector field on $\cU$ defined by
\beq \label{Xv}
X_\vv \= \vv^0\frac{\partial}{\partial t} + \vv^i\frac{\partial}{\partial y^i}
+ \sum_{a=1}^{k} \vv^i_{(a)}\frac{\partial}{\partial y^i_{(a)}}
\eeq
where
\beq \label{Xvbis}
\vv^i_{(a)} \= \frac{d^a}{dt^a}\left(\vv^i-y^i_{(1)}\vv^0\right)+y^i_{(a+1)}\vv^0\, .
\eeq
(in this formula, we  assume $y^i_{(k+1)} = 0$).
Notice that, by \eqref{Xv} and \eqref{Xvbis}, we may also write $X_\vv$  as
 \beq \label{Xvter}
X_\vv \= \vv^0\frac{d}{d t} + \left(\vv^i- y^i_{(1)} \vv^0\right)\frac{\partial}{\partial y^i}
+ \sum_{a=1}^k  \frac{d^a}{dt^a}\left(\vv^i - y^i_{(1)} \vv^0\right)\frac{\partial}{\partial y^i_{(a)}}
\eeq
\medskip
\begin{prop} \label{lemma61} If $\dim M \geq 2$,
a vector field $X$ on $\cU$ is a $\cD$-symmetry if and only if $X=X_\vv$ for some  $\vv = (\vv^i)$
such that $\displaystyle\frac{\partial \vv}{\partial y^i_{(k)}}  = 0$ for all $1 \leq i \leq n$.
\end{prop}

\begin{pf}
By Proposition \ref{equiv-Dsimm}, a vector field
$$
X = X^0\frac{\partial}{\partial t} + X^i\frac{\partial}{\partial y^i}
+ X^i_{(1)}\frac{\partial}{\partial y^i_{(1)}} + \ldots + X^i_{(k)}\frac{\partial}{\partial y^i_{(k)}}
$$
on $\cU$ is a $\cD$-symmetry if and only if it satisfies the equations
\beq \label{6.2} \omega^i_{(a)}\left(\cL_X \frac{d}{dt}\right) = 0\, , \qquad
\omega^i_{(a)}\left(\cL_X \frac{\partial}{\partial y^j_{(k)}}\right) = 0 \eeq
for any $0 \leq a \leq k-1$. We recall that
$$
\cL_X \frac{d}{dt} =  - \frac{d X^0}{dt}\frac{\partial}{\partial t} + \sum_{a = 0}^{k-1}\left(X^i_{(a+1)}
- \frac{d X^i_{(a)}}{dt}\right)\frac{\partial}{\partial y^i_{(a)}} - \frac{d X^i_{(k)}}{dt} \frac{\partial}{\partial y^i_{(k)}}\ .
$$
Hence, the first set of equations in \eqref{6.2} means that, for any $0 \leq a \leq k-1$,
\beq \label{Omega=0}
0=\omega^i_{(a)}\left(\cL_X \frac{d}{dt} \right) = X^i_{(a+1)} - \frac{d X^i_{(a)}}{dt} + y^i_{(a+1)} \frac{d X^0}{dt}\ .
\eeq
This shows that all components $X^i_{(a)}$, $a \geq 1$, are uniquely determined by the components $X^i$
and, by induction, one can check that $X$ is as in \eqref{Xv}.
\par

In order to conclude, it suffices to show that the other equations in \eqref{6.2} are equivalent to
\beq \label{6.5} \frac{\partial X^0}{\partial y^j_{(k)}} =  \frac{\partial X^i}{\partial y^j_{(k)}}  =
\frac{\partial X^i_{(a)}}{\partial y^j_{(k)}} = 0\eeq
for any $0 \leq a \leq k-1$, so that $\displaystyle\frac{\partial \vv^\ell}{\partial y^i_{(k)}}  =
\frac{\partial X^\ell}{\partial y^i_{(k)}}  = 0$.
Indeed, denoting by $z^A$ an arbitrary coordinate amongst $(t, y^i, y^i_{(a)})$, one has that
$\cL_X \frac{\partial}{\partial y^j_{(k)}} = -  \frac{\partial X^A}{\partial y^j_{(k)}} \frac{\partial}{\partial z^A}.$
This means that the second set of equations in \eqref{6.2} is equivalent to
\beq \label{6.4}  \frac{\partial X^i_{(a)}}{\partial y^j_{(k)}} =  y^i_{(a+1)} \frac{\partial X^0}{\partial y^j_{(k)}}\ ,
\quad 0 \leq a \leq k-1.\eeq
Now, setting $a=k-1$ and taking the derivative of \eqref{6.4} w.r.t. $y^i_{(k)}$ for some
$i \neq j$, we get
$$
\frac{\partial^2 X^i_{(k-1)}}{\partial y^i_{(k)} \partial y^j_{(k)}} = \frac{\partial X^0}{\partial y^j_{(k)}} +
y^i_{(k)} \frac{\partial^2 X^0}{\partial y^i_{(k)}\partial y^j_{(k)}}\, .
$$
On the other hand, considering equation \eqref{6.4} with $j=i$ and taking the derivative w.r.t. $y^j_{(k)}$ we have
$$
\frac{\partial^2 X^i_{(k-1)}}{\partial y^j_{(k)} \partial y^i_{(k)}} =
y^i_{(k)} \frac{\partial^2 X^0}{\partial y^j_{(k)}\partial y^i_{(k)}} \, .
$$
Taking the difference, we obtain $\frac{\partial X^0}{\partial y^j_{(k)}} = 0$.
Inserting this in \eqref{6.4}, equalities \eqref{6.5} follow.
\end{pf}
We have now all the ingredients for the two parts of the Noether Theorem, which are stated and proved
in the next section.\par
\bigskip

\subsection{Noether Theorem}
\begin{definition}\label{formaradicale}
Let $[\alpha]$ be a variational class of 1-forms on $J^k(E)$, determined by a 1-form $\a$, which is locally variationally equivalent to $1$-forms $L dt$ of order $r$ for some $r \leq \frac{k}{2}$. A 1-form $\a_o \in [\a]$ is called {\it of Poincar\'e-Cartan type} if $d \a_o$ is a source form modulo a holonomic 2-form.
\end{definition}
The main example  of such  kind of  1-forms  is  given by the Poincar\'e-Cartan form $\a_o = p_i dq^i - H dt$ discussed in (4) of \S \ref{sec2.3}. In fact
$$d\a_o =  \frac{\partial H}{\partial p_j}Êd p_j \wedge dt + \frac{\partial H}{\partial q^k}Êd q^k \wedge dt  + d p_i \wedge d q^i\ ,$$
which is a source form on any jet space $J^k(E)$, $k \geq 1$,  of  the trivial bundle $\pi: E = T^* \bR^n \times \bR\longrightarrow \bR$.\par
\medskip
Note  that if  $\a$ is a 1-form on $J^k(E)$,  satisfying the assumptions of \eqref{formaradicale}, then    for  any $u \in J^k(E)$  there exists a neighborhood $\cU$ of $u$ such that  the variational class  $[\a|_{\cU}]$  contains a  1-form of Poincar\'e-Cartan type.  This can be directly seen as follows:  consider a neighbourhood $\cU$ admitting  a system of adapted coordinates, and let  $\s \in [d\a|_{\cU}]$ be the source form  described  in \eqref{2.12}.  Then $\s = d \a|_{\cU}+ d \mu + \l = d(\a|_{\cU}+ \mu) + \l$,  for some holonomic  $\mu$ and $\l$,  and   $\a_o = \a|_{\cU} + \mu$ is a 1-form of Poincar\'e-Cartan type in the variational class  $[\a|_{\cU}]$.\par
\smallskip
We also remark that,  replacing $J^k(E)$ by a jet space of higher order,  one may safely assume  that  the variational class $[\a|_{\cU}]$ contains at least one   1-form of Poincar\'e-Cartan type {\it of order $r \leq k-1$}.  We will shortly see that such harmless  assumption is  often quite convenient.
\par
\medskip
The notion of 1-forms of Poincar\'e-Cartan type leads to the following useful characterisation of infinitesimal symmetries of a given action.
As in Proposition \ref{lemma61},   we consider as fixed    a system of adapted coordinates
$\wh\xi^{(k)} = (t, y^i,y^i_{(a)})$ on an open subset  $\cU \subset J^k(E)$ and  for any $\bR^{n+1}$-valued smooth map
$\vv = (\vv^i)$ on $\cU$,  we denote by $X_\vv$ the associated vector field  defined in \eqref{Xv}.
\bigskip
\begin{prop} \label{lemma62} Assume that $\dim M \geq 2$ and let $\a_o$ be a 1-form of Poincar\'e-Cartan type in $[\a]$ of order $r \leq k-1$ and $
X_\vv$ a $\cD$-symmetry on $\cU$ associated with  $\vv = (\vv^i)$.
Then $X_\vv$ is an infinitesimal symmetry for $\cI_{[\a]}$ if and only if  it  satisfies the linear differential equation
\beq  \label{3.25bis} \frac{d}{dt}(\a_o(X_\vv))  =   \s\left(\frac{d}{dt}, X_\vv\right) \ ,\eeq
where $\s$  is any   source form    of the variational class $[\a|_{\cU}]$.
\end{prop}
\begin{pf} Let  $\l$ be the   holonomic 2-form  defined by  $\l = d \a_o - \s $.
By Proposition \ref{equiv-Dsimm} (2) and the fact that
$$\imath_{\frac{d}{dt}} \l = \imath_{\frac{\partial}{\partial y^j_{(k)}}} \l = \imath_{\frac{\partial}{\partial y^j_{(k)}}} \s = 0\ ,$$
   $X_\vv$  is an infinitesimal symmetry for $\cI_{[\a]}$ if and only if
  \beq \label{3.26}
\left\{\begin{array}{l} \cL_{X_\vv} \a_o \left(\frac{d}{dt}\right)  = d(\a_o (X_\vv)) \left(\frac{d}{dt}\right) + d\a_o(X_\vv, \frac{d}{dt}) =\\
\ \\
\phantom{aaaaaaaaa} = \frac{d}{dt}(\a_o(X_\vv)) +   \s(X_\vv, \frac{d}{dt}) = 0\ , \\
 \\
 \cL_{X_\vv} \a_o \left( \frac{\partial}{\partial y^i_{(k)}}\right) = \frac{\partial \a_o(X_\vv)}{\partial y^i_{(k)}} +   \s(X_\vv, \frac{\partial}{\partial y^i_{(k)}}) =  \frac{\partial  \a_o(X_\vv)}{\partial y^i_{(k)}}  = 0 \ .
 \end{array}\right.
 \eeq
 Since   $\frac{\partial X^A_\vv}{\partial y^i_{(k)}}  = 0$ for all components $X^A_\vv$ of $X_\vv$ (Proposition \ref{lemma61}) and $\a_o$ is of order $r \leq k-1$, the second equality is trivially satisfied for any $1 \leq i \leq k$.
By the  first equation in  \eqref{3.26},   the claim follows.
\end{pf}
We can now state and prove the Noether Theorem in its two parts, direct and inverse.

\begin{theo}[\bf Noether Theorem -- first part]\label{directNoether}
Let $[\alpha]$ be a variational class of 1-forms on $J^k(E)$ and  assume that $\a_o$ is a 1-form of Poincar\'e-Cartan type in $[\a]$.  \par
 If a vector field $X$ on $\cU \subset J^k(E)$  is   an infinitesimal symmetry for $\cI_{[\a]}$ (considered as functional  on  sections in $\cU$), then
 $$f^{(X)} \= \imath_X \a_o: \cU \longrightarrow \bR$$
 is a constant of motion for the  variational principle of   $\cI_{[\a]}$.
\end{theo}
\begin{pf} By definition of 1-forms of Poincar\'e-Cartan type,  $d \a_o = \s + \l$, where $\s$ is a source form in $[d\a_o]$ and $\l$ is a holonomic 2-form.
It follows that, for any section  $\g: I \longrightarrow E$
$$\frac{d f^{(X)} \circ \g^{(k)}}{dt} = d(\imath_X \a_o)\left(\frac{d \g^{(k)}}{dt} \right) = $$
$$ = \cL_X \a_o\left(\frac{d \g^{(k)}}{dt} \right)  -
d \a_o\left(X, \frac{d \g^{(k)}}{dt} \right) \overset{\cL_X \a_o\ \text{is holonomic}}= -d \a_o\left(X, \frac{d \g^{(k)}}{dt} \right) =$$
$$ = - \s\left(X, \frac{d \g^{(k)}}{dt} \right) \ .$$
Since $\s$ is a source form of $[d \a_o]$, by  Theorem \ref{cor2.3.4}, if $\g$  is a solution of the variational principle of $\cI_{[\a]}$, we have
$\frac{d f^{(X)} \circ \g^{(k)}}{dt} =  - \s\left(X, \frac{d \g^{(k)}}{dt} \right)  = 0$.
\end{pf}
\bigskip
Now, in order to state and prove the inverse of this result,  we  need to consider  a new  notion. \par
\smallskip
Let $[\alpha]$ be a variational class of 1-forms  of  $J^k(E)$
and assume that $\s = \s_i \o^i_{(0)} \wedge dt$  is a source form  of the kind    \eqref{2.12}  on some open subset $\cW \subset J^k(E)$. Assume also that $\s$ is
 of order $r_o \leq k-1$ and   consider  the differentials $d\s_i$  of the components $\s_i$ of $\s$. By the assumption on the order of $\s$,
 such differentials are equal to
$$d \s_i = \frac{\partial \s_i}{\partial t} dt +  \sum_{a = 0}^{k-1}\frac{\partial \s_i}{\partial y^j_{(a)}} dy^j_{(a)} =  \frac{d \s_i}{dt} d t +  \sum_{a = 0}^{k-1}\frac{\partial \s_i}{\partial y^j_{(a)}} \o^j_{(a)}\ .$$
Due to this, for any $k$-th order lift  $\g^{(k)}: I \longrightarrow \cW$   of a section $\g$ of $E$,  we have
$$d\left(\s_i(\g^{(k)}(t))\right) =  d \s_i\left(\dot \g^{(k)}(t)\right) = \left.\frac{d \s_i}{dt} \right|_{\g^{(k)}(t)}\ .$$
Hence a lifted section $\g^{(k)}$ corresponds to a solution of the Euler-Lagrange  equations
\beq \label{firstsystem} \s_i(\g^{(k)}_t) = 0\ ,\qquad 1 \leq i \leq n\ ,\eeq
 if and only if it is a solution of the system of partial differential equations
\beq \label{firstprol} \s_i(\g^{(k)}_t) =  \frac{d \s_i}{dt}(\g^{(k)}_t)= 0\ , \qquad 1 \leq i \leq n\ .\eeq
The system   \eqref{firstprol} is usually called {\it first  prolongation} of  \eqref{firstsystem}. We stress the fact  {\it if the functions (i.e.,  $0$-forms) $\s_i$ which defined the Euler-Lagrange equations are of order $r_o$,  the  functions   that define the first prolongation \eqref{firstprol} are  $0$-forms of order  $r_o + 1 \leq k$}. \par
\smallskip
 Consider  now   the integer $p_o \= k-r_o$. Iterating the above argument,  we can directly prove  that the system \eqref{firstsystem} is equivalent to
\beq \label{pthprol} \s_i(\g^{(k)}_t) =  \frac{d \s_i}{dt}(\g^{(k)}_t)= \ldots =  \left(\left(\frac{d}{dt}\right)^{p_o}\!\!\!\!\!\!\s_i\right)(\g^{(k)}_t) = 0 \ ,\qquad 1 \leq i \leq n\ .\eeq
 We  call it    {\it  full  prolongation of
\eqref{firstsystem}} on the $k$-order jet space $J^k(E)$. \par
\medskip
Note that  {\it the order of the collection of functions  appearing in a  full prolongation  is  generically not less than $k$}.
\par
\bigskip
\begin{definition} Let $F_\s: \cW \subset J^k(E)\longrightarrow \bR^{n \cdot (p_o+1)}$ be the smooth function
\beq \label{effesse} F_\s \= \left(\s_{i}, \frac{d}{dt} (\s_{j}),  \left(\frac{d}{dt}\right)^2 (\s_{\ell}), \ldots,  \left(\frac{d}{dt}\right)^{p_o} (\s_{m})\right)\eeq
and set  $Z_\s  \= \{\ u \in \cW\ : \ F_\s(u) = 0\ \} \subset J^k(E)$. The system of Euler-Lagrange equations \eqref{firstsystem} is  called {\it   regular in $J^k(E)$} if  the map $F_\s$ is a submersion  at all points of $Z_\s$
\end{definition}
We may now state the second part of Noether Theorem.
\begin{theo}[\bf Noether Theorem -- second part] \label{inverseNoether} Assume that $\dim M \geq 2$ and let $\a \in [\a]$ be a 1-form of Poincar\'e\--Cartan type of order $r \leq k-1$ in  a variational class
 $[\alpha]$  of 1-forms  on  $J^k(E)$.
Assume also that  there exists an open subset $\cW \subset J^k(E)$, where the  following non-degeneracy conditions are satisfied:
\begin{itemize}
\item[a)]    there exists a source form $\s = \s_i \o^i_{(0)} \wedge dt$  of order $r_o \leq k-1$ on $\cW$ of the kind   \eqref{2.12},     which determines a   system of Euler-Lagrange equations $\s_i = 0$, which is regular in $J^k(E)$;
\item[b)] $\left.\a_o\left(\frac{d}{dt}\right)\right|_{u}  \neq 0$ at all $u$'s in $\cW$.
\end{itemize}
 \par
If $f:   \cW \longrightarrow \bR$ is a constant of motion of order $k-1$ for the  variational principle of   $\cI_{[\a]}$,  then there exist
\begin{itemize}
\item[1)] a neighborhood $\cU$ of $Z_\s  = \{\ u \in \cW\ : \ F_\s(u) = 0\ \}$, where $F_\s$ is defined in \eqref{effesse};
\item[2)]   an infinitesimal symmetry  $X^{(f)}$ for $\cI_{[\a]}$ on $\cU$;
\item[3)] a $p_o$-tuple of constants of motion $(g^{(1)}, \ldots, g^{(p_o)})$, $p_o = k - r_o -1$,  on $\cU$, vanishing  at all points $\g^{(k)}(t)$ of all lifts of the solutions of the variational principle
\end{itemize}
such  that
\beq \label{noether2}  \imath_{X^{(f)}} \a = f|_{\cU} + g^{(1)} + \ldots + g^{(p_o)}\  .\eeq
\end{theo}
\begin{pf} Consider   a  system of adapted coordinates  $\wh\xi^{(k)} = (t, y^i,y^i_{(a)})$ and let $\s = \s_i \o^i_{(0)} \wedge dt $ on $\cW$
be a source form satisfying the non-degeneracy condition (a).  By Propositions \ref{lemma61} and  \ref{lemma62},  we  need  to show that there exists a neighbourhood $\cU$ of $Z_\s$, a smooth $\bR^{n+1}$-valued map $\vv = (\vv^0, \vv^i): \cU \to \bR^{n+1}$ and  $p_o$ constants of motion $g^{(i)}$ on $\cU$,   vanishing on lifts    $\g^{(k)}(I)$ of solutions, such that the  vector  field $X_\vv$ satisfies  the system of linear equations
\beq\label{ahah}  \a(X_\vv) = f + \sum_{\ell = 1}^{p_o} g^{(\ell)} \ ,\qquad  (\imath_{\frac{d}{dt}} \s) (X_\vv)  = \frac{d f}{dt} + \sum_{\ell = 1}^{p_o}\frac{d g^{(\ell)}}{dt} \ .
\eeq
If we  express $\a$ and $\s$ as sums of the form
$$\a = \a_0 dt  + \sum_{\smallmatrix 0 \leq a \leq k-1\\ 1\leq i  \leq n \endsmallmatrix} \a^{(a)}_i \o^i_{(a)}\ ,\qquad \s = \sum_{1\leq i  \leq n}\s_i \o^i_{(0)} \wedge dt \ ,$$
  equations \eqref{ahah} become
\beq \label{3.29} \left\{\begin{array}{l}  \displaystyle  \vv^0 \a_0 = -  \sum_{i =1}^n\left(\vv^i - y^i_{(1)}\vv^0\right) \a_i^{(0)} - \!\!\!  \sum_{\smallmatrix 1 \leq a \leq k-1\\ 1\leq i  \leq n \endsmallmatrix}\!\!\!\!
\frac{d^a \left(\vv^i - y^i_{(1)}\vv^0\right)}{dt^a} \a^{(a)}_i    +\ \ \ \ \ \\
\hfill +  f  + \displaystyle \sum_{\ell = 1}^{p_o} g^{(\ell)} \ ,\\
\ \\
\displaystyle \sum_{i = 1}^n \left(\vv^i - y^i_{(1)} \vv^0\right)\s_i = - \frac{d f}{dt} - \sum_{\ell = 1}^{p_o}\frac{d g^{(\ell)}}{dt}  \ . \end{array}\right.\eeq
We claim that the function $\frac{d f}{dt}: \cW \longrightarrow \bR$ vanishes identically on $Z_\s$. Indeed,
since  $Z_\s = \{F_\s = 0\}$ is equal to the collection of the jets of  the ($k$-th order lifts of) solutions to the variational principle,  for any $u \in Z_\s$,
$$\left.\frac{d}{dt}\right|_u -  \dot \g^{(k)}|_{t_o} \in \text{Span}\left\{\left.\frac{\partial}{\partial y^i_{(k)}}\right|_u\right\}\ ,$$
where we denoted by $\g^{(k)}$ the $k$-th order lift of a  solution with $u = \g^{(k)}(t_o)$.
Since $f$ is a   constant of motion and it  is   of order $k-1$, we get
$$df\left(\left.\frac{d}{dt}\right|_u\right) = df\left(\dot \g^{(k)}(t_o)\right)  = 0\ , $$
which  proves the claim.\par
\medskip
From this, the  fact that   $F_\s: \cW \longrightarrow \bR^{n (p_o+1)}$ is a submersion at any  $u \in Z_\s$ and   standard properties of  submanifolds (see e.g., \cite{Mi}, Lemma 2.1 and \cite{Ol}, Prop. 2.10),
there exists a neighborhood $\cU \subset \cW$ and   $n\cdot (p_o+1)$ smooth functions $\wh \vv^j_{(\ell)}$,  $1 \leq j \leq n$,  $0 \leq \ell \leq p_o$,  on $\cU$ (not uniquely determined!), such that
\beq\label{acci}   - \frac{d f}{dt} =  \sum_{i = 1}^n \wh \vv^i_{(0)} \s_i +  \sum_{i = 1}^n \wh \vv^i_{(1)} \frac{d \s_i}{dt} + \ldots + \sum_{i = 1}^n \wh \vv^i_{(p_o)} \left(\frac{d}{dt}\right)^{p_o}\!\!\!(\s_i)\ .\eeq
Let $g^{(1)}: \cU \to \bR$ be the smooth function defined by
\beq g^{(1)} \= \sum_{i = 1}^n\wh \vv^i_{(p_o)} \left(\frac{d}{dt}\right)^{p_o-1}\!\!\!(\s_i)\ .\eeq
This function  vanishes identically on the jets of the solutions  (it is pointwise equal to a linear combination  components of the map  $F_\s $) and it is therefore a constant of motion. Furthermore,
$$\sum_{i = 1}^n \wh \vv^i_{(p_o)} \left(\frac{d}{dt}\right)^{p_o}\!\!\!(\s_i) = \frac{d g^{(1)}}{dt} - \sum_{i = 1}^n \frac{d}{dt}( \wh \vv^i_{(p_o)})\cdot \left(\frac{d}{dt}\right)^{p_o-1}\!\!\!\!\!\!(\s_i) \ ,$$
so that  \eqref{acci} can be re-written in the form
\beq\label{accibis}    - \frac{d f}{dt} - \frac{d g^{(1)}}{dt} = \sum_{a = 0}^{p_o -1} \left(\sum_{i = 1}^n \wh \vv^i_{(a)} \left(\frac{d }{dt}\right)^a\!\!\!\!\!(\s_i)\right) +  \sum_{i = 1}^n \wt \vv^i_{(p_o-1)} \left(\frac{d}{dt}\right)^{p_o-1}\!\!\!\!\!\!(\s_i)  \eeq
where we set
$$ \wt \vv^i_{(p_o -1)}\= \wh \vv^i_{(p_o-1)} -  \frac{d}{dt}(\wh \vv^i_{(p_o)}) \ .$$
Iterating this line of arguments, we conclude  that \eqref{acci} is equivalent to an equality of the form
\beq\label{acciter}  - \frac{d f}{dt} - \frac{d g^{(1)}}{dt} - \ldots -\frac{d g^{(p_o)}}{dt} =  \sum_{i = 1}^n \wt \vv^i \s_i \ ,\eeq
for some appropriate smooth functions $\wt \vv^i, g^{(\ell)}: \cU \longrightarrow \bR$, where the $g^{(\ell)}$  are  constants of motion that vanish identically on the jets of the  solutions of the variational principle. \par
\medskip
Since $\a_0 = \a\left(\frac{d}{dt}\right)$ is nowhere vanishing on $\cW$,  we may consider the   function
$$\vv^0 \= -   \sum_{i =1}^n\wt \vv^i  \frac{\a_i^{(0)}}{\a^0} -  \sum_{\smallmatrix 1 \leq a \leq k-1\\ 1\leq i  \leq n \endsmallmatrix}\!\!\!\!
\frac{d^a \wt \vv^i }{dt^a} \frac{\a^{(a)}_i}{\a_0}   +  \frac{f + \sum_{\ell = 1}^{p_o}g^{(\ell)} }{\a_0}$$
and   the corresponding  $(n+1)$-tuple of functions on $\cU \subset \cW$
$$\vv \= \left(\ \ \vv^0, \  \vv^1= \wt \vv^1 + y^1_{(1)} \vv^0, \quad \ldots \quad ,\  \vv^n= \wt \vv^n + y^n_{(1)} \vv^0\ \ \right)\ .$$
By construction, $\vv$ satisfies \eqref{3.29} and  $X^{(f)} \= X_\vv$  is an infinitesimal symmetry satisfying \eqref{noether2}.
\end{pf}
\bigskip
\subsection{Correspondence between infinitesimal symmetries and constants of motion}
\label{section33}
Let $[\a]$ be a variational class on $J^k(E)$, which is locally determined by a 1-form $L dt$  of order $r$ with $2r \leq k$,  and assume that $X$  is an infinitesimal symmetry
$X$ for the action $\cI_{[\a]}$ on some open subset $\cU \subset J^k(E)$.  \par
By the proof of the first part of the Noether Theorem,
if  $\a_o$,  $\a'_o$ are   distinct 1-forms of Poincar\'e-Cartan type in $[\a]$ and $f^{(X)}$ and $g^{(X)}$ are the constants of motion associated with $X$ via $\a_o$ and $\a'_o$, i.e.,
$$f^{(X)}\=  \imath_X \a_o \ ,\qquad  g^{(X)} \= \imath_X \a'_o\ ,$$
the difference $h = f^{(X)} - g^{(X)}$ is   a constant of motion with the property that,  for any $k$-lift $\g^{(\k)}$ of a section of $E$  (here, $\s$ is a source form of $[d \a_o]$)
$$\frac{d h \circ \g^{(k)}}{dt}  = \s\left(X,  \frac{d \g^{(k)}}{dt}\right) - \s\left(X,  \frac{d \g^{(k)}}{dt}\right)  = 0\ .$$
\medskip
It is  therefore convenient to consider the following definition.
\begin{definition}  An infinitesimal symmetry
$X$ for the action $\cI_{[\a]}$ on   $\cU \subset J^k(E)$  is called
\begin{itemize}
\item[1)] {\it $\a_o$-trivial} if the constant of motion $f^{(X)} \= \imath_X \a_o $, determined by   $\a_o \in [\a]$ of Poincar\'e-Cartan type, is  the zero function $f^{(X)} = 0$;
\item[2)] {\it trivial} if  for any $u \in \cU$ there exists at least one  $\a_o \in [\a]$ of Poincar\'e-Cartan type on a neighbourhood $\cU' \subset \cU$ of $u$ such that  $f^{(X)} \= \imath_X \a_o $ is   constant  on any  $k$-th order lift $\g^{(k)}$ of  a section $\g$ of $E$.
\end{itemize}
\end{definition}
By previous observations, the property of being trivial {\it does not} depend on the  choice of the  1-form $\a_o$  of Poincar\'e-Cartan type and it is equivalent to the condition
$$\s \left(X, \frac{d}{dt}\right) = 0\ ,$$
where $\s$ is an arbitrary source form  $[d\a_o]$.\par
\medskip
Take now a fixed variational class $[\a]$  on $J^k(E)$, with the usual  assumption that $\a \simeq L dt$ for some 1-form $L dt$ of order $r$ with $2r \leq k$,  and let  $\a_o \in [\a]$ be of  Poincar\'e-Cartan type of order $r' \leq k-1$ on  some open set $\cU\subset J^k(E)$. Fix $u_o \in \cU$ and consider the following classes of germs at $u_o$ (here,  given a vector field $X$ or a function $f$, we denote  by $\underline X$ and $\underline f$, respectively, their germs at $u_o$):
$$\begin{array}{l}
\Symm \=  \{\text{germs at}\ u_o \ \text{of infinitesimal symmetries  of}\ \cI_{[\a]}\ \}\\
\ \\
\Triv^{(\a_o)}\= \{\ \underline X \in \Symm\ :\  X\ \text{is an}\ \a_o\text{-trivial symmetry}\ \ \} \\
\ \\
\Triv\= \{\ \underline X \in \Symm\ :\  X\ \text{is a}\ \text{trivial symmetry}\ \ \} \ \\
\ \\
\First\= \{\ \text{germs at}\ u_o \ \text{of constants of motion  for}\ \cI_{[\a]}\  \} \\
\ \\
\Null\= \{\  \underline f \in \First\ :\ f\ \text{is identically equal to $0$  at  points of  solutions}\  \} \\
\ \\
\Const\= \{\ \underline f \in \First\ :\ f\ \text{is constant along any section}\ \g^{(k)}\  \} \\
\ \\
\end{array}
$$
{\it All such classes of germs have  natural structures of vector spaces. The  space $\Symm$  is   also endowed with a natural Lie algebra structure, given by the usual Lie brackets between vector fields. }\par
\medskip
Using  the above  notation, when $\dim M \geq 2$ and the non-degeneracy conditions (a) and (b) of Theorem \ref{inverseNoether} are satisfied,  the two parts of Noether Theorem can be restated saying that {\it for any given choice of  a  1-form $\a_o \in [\a]$ of Poincar\'e-Cartan type of order $r' \leq k-1$, there exists a natural  {\rm surjective} linear map}
\beq \label{noetherter} \varphi^{(\a_o)}: \Symm \longrightarrow \First/\Null\ ,\eeq
\par
 From the definition of the map $\varphi^{(\a_o)}$,  one has that   $\ker \varphi^{(\a_o)} = \Triv^{(\a_o)}$ and the above homomorphism induces an  isomorphism of vector spaces
$$ \imath^{(\a_o)}: \Symm/\Triv^{(\a_o)} \overset{\sim}\longrightarrow \First/\Null\  . $$
\par
This isomorphism  {\it does depend} on the choice of $\a_o$. However, if one considers
the  quotients of the vector  spaces $\Symm$ and   $\First$ by the subspaces $\Triv$ and  $\Null + \Const$, respectively,   the surjective  map \eqref{noetherter}
establishes  a vector space isomorphism
$$ \imath: \Symm/\Triv \overset{\sim}\longrightarrow \First/(\Null + \Const)\  , $$
which is  now {\it   independent on the choice of $\a_o$}.\par
\medskip
A priori, there is no  reason for $\Triv^{(\a_o)}$ or $\Triv$ to be  ideals of the Lie algebra $\Symm$. Due to this, {\it the quotients $\Symm/\Triv^{(\a_o)}$ and $\Symm/\Triv$ cannot be expected to have a natural Lie algebra structure}.  \par
\medskip
However, something can be said on this regard, provided that  we
consider the following restricted class of infinitesimal symmetries. \par
\begin{definition} Given $\a_o \in [\a]$ of Poincar\'e-Cartan type and  with the above conditions satisfied, an infinitesimal symmetry $X$ for $\cI_{[\a]}$  is called
{\it $\a_o$-symmetry } if
$\cL_X \a_o = 0$.
\end{definition}
Denote by $\Symm^{(\a_o)} \subset \Symm$ the subalgebra of the germs at $u_o$ of $\a_o$-symmetries.  We claim that the  Lie brackets between vector fields  induce a linear action  of  $\Symm^{(\a_o)}$ on $ \Triv^{(\a_o)} $. Indeed, if $\underline X \in \Symm^{(\a_o)}$ and
$\underline Y \in \Triv^{(\a_o)}  $
$$\imath_{[X, Y]} \a_o = d (\a_o(Y))(X) - \cL_{X} \a_o(Y)= d (\a_o(Y))(X) \overset{\a_o(Y) \equiv 0}= 0$$
showing that the germ  $\underline{[X, Y]}$ is in $\Triv^{(\a_o)}$. Hence,  the  map
\beq \wt \ad: \Symm^{(\a_o)} \longrightarrow \Hom\left(\frac{\Symm}{\Triv^{(\a_o)}}, \frac{\Symm}{\Triv^{(\a_o)}}\right)\ ,\ $$
$$  \wt \ad_{\underline X}(\underline Z \!\!\! \mod \Triv^{(\a_o)}) \= \underline{[X, Z]} \mod \Triv^{(\a_o)}\ .\eeq
is well-defined and is a linear representation.  Composing  with the isomorphism $\imath^{(\a_o)}$, we get the following linear map for any $X \in \Symm^{(\a_o)}$:
$$ \r(X) : \First\longrightarrow \First\ ,\ $$
$$ \r(\underline X)(\underline f) \= \underline{\imath_{[X,Z^{(f)}]} \a_o}  = \underline{X(\imath_{Z^{(f)}} \a_o)} - \underline{\imath_{Z^{(f)}} \cL_X \a_o} = \underline{X(f)} ,$$
where   $\underline Z^{(f)}$ is any  germ in $\Symm^{(\a_o)}$ that  is mapped onto $f$ by  $\varphi^{(\a_o)}$. By construction, the map $\r$ determines a linear representation of $\Symm^{(\a_o)}$ and we  have  the following:
\begin{prop} Given  $u_o \in J^k(E)$ and  $\a_o \in [\a]$   of Poincar\'e-Cartan type and satisfying the hypothesis of Theorem \ref{inverseNoether}, the map
\beq  \label{repres1}\r: \Symm^{(\a_o)} \longrightarrow \Hom\left(\First/\Null, \First/\Null\right)\ , $$
$$ \r(\underline X)([f]_\Null)   \= \left[\underline{X(f)}\right]_\Null\ \eeq
is  a linear representation of  the space of (germs of) $\a_o$-symmetries $\Symm^{(\a_o)}$ on the quotient space of (germs of) constants of motion  $\First/\Null$.
 \end{prop}
\par
\medskip
\begin{rem}A similar argument can be used to show  the existence of a natural linear representation of  $\Symm^{(\a_o)}$ also on the quotient space $\First/(\Null + \Const)$.
\end{rem}
\bigskip
\section{Infinitesimal symmetries and  Hamiltonian vector fields in Hamiltonian mechanics}
\setcounter{equation}{0}
\subsection{Notational issues}
From now on, we assume that the  configuration space  $M$  is a cotangent bundle $M = T^* N$ of an $n$-dimensional manifold $N$. \par
\medskip
We  denote by $\wh \pi: T^*N \to N$ the canonical projection of $T^*N$  and
for any system of coordinates $\h = (q^1, \ldots, q^n): \cU \subset N \longrightarrow \bR^n$ of $N$, we call {\it associated coordinates on $T^*N$} the map
$$\xi_\h: \wh \pi^{-1}(\cU) \subset T^* N \longrightarrow \bR^{2n}\ ,$$
which associates to any 1-form $\b = p_i d q^i|_{x} \in T^*_x N$,  $\h(x) = (q^1, \ldots, q^n)$,    the  coordinates
$$\b = p_i d q^i|_{x} \overset{\xi_\h} \longrightarrow= (q^1, \ldots, q^n, p_1, \ldots, p_n)\ .$$
\par
\medskip
In  the following, we  consider only this kind of coordinates  on $T^*N$   and  the systems of adapted
coordinates  on $J^k(E)$, $E = T^*N \times \bR$, are  assumed  associated with such coordinates and of the form
$$\wh \xi^{(k)} = (t, q^i, p_j, q^i_{(1)}, p_{j(1)}, \ldots,q^i_{(k)}, p_{j(k)}): \cU \subset J^k(E) \longrightarrow \bR^{2n(k+1)+1}\ .$$
\par
\bigskip
The components of a vector field $X$ on  $\cU \subset J^k(E)$ along the coordinate vector fields   $\frac{\partial}{\partial q^i_{(a)}}$ (resp. $\frac{\partial}{\partial p_{j(a)}}$) are denoted by $X^i_{(a)}$ (resp. $X_{j(a)}$), that is $$X = X^0 \frac{\partial}{\partial t} +  X^i \frac{\partial}{\partial q^i} +  X_j \frac{\partial}{\partial p_j} +  X^i_{(1)} \frac{\partial}{\partial q^i_{(1)}} + X_{j(1)} \frac{\partial}{\partial p_{j(1)}} + \ldots\ \ . $$
\par
\medskip
The holonomic 1-forms \eqref{holonomicforms} are now denoted by
$$\o^i_{(a)} \= dq^i - q^i_{(a)}dt \ ,\qquad \o_{i(a)} \= d p_i - p_{i(a)} dt\ .$$
\par
\medskip
We finally  denote by   $\q$  and $\O$ the {\it tautological 1-form} and {\it canonical symplectic 2-form}, respectively, of $T^*N$. We recall that they are defined by
$\q|_\b\= \b(\wh \pi_*(\cdot))$ and $\O = d \q$
and that,  in   coordinates   $\xi_\h = (q^i, p_j)$, they are given by  the well-known expressions
$$\q = p_i d q^i\ ,\qquad \O = d\q = d p_i \wedge d q^i\ .$$
\par
\bigskip
\subsection{Infinitesimal symmetries of Hamiltonian actions}
According to the standard  terminology  of Hamiltonian mechanics, a {\it (time independent) Hamiltonian} is a smooth real function
$H: \cU \subset M = T^*N \longrightarrow \bR$ defined on some open subset of $T^*N$. \par
\smallskip
For a given Hamiltonian $H$, let us consider the following definition.
\begin{definition}
 The {\it   Poincar\'e-Cartan 1-form of $H$} is the 1-form $\a^H$ on the bundle  $\pi: E = \cU \times \bR \subset T^*N \times \bR \longrightarrow \bR$ defined by
$$\a^H \= \q - H dt $$
$$(\ \ \text{in coordinates,} \  \a^H \= p_i dq^i - H dt \ )\ .$$
 For any $k \geq 1$,  the {\it Hamiltonian action  of $H$ on $J^k(E)$} is the action $\cI_{[\a^H]}$, defined  by  the variational class on $J^k(E)$ of  the (pull-back on $J^k(E)$ of)
$\a^H$.
\end{definition}
\par
\medskip
If  $\a^H$ is considered as a  1-form of $J^1(E)$, we may  see that it is  (locally) variationally equivalent to the 1-form
$$\a^H - p_i \o^i =   (H - p_i q^i_{(1)}) dt  \ .$$
This means that  the action $\cI_{[\a^H]}$   is (locally)   determined by the  Lagrangian  $L= H - p_i q^i_{(1)}$, which is clearly  of order $1$. Furthermore,
$$d \a^H = \O - dH \wedge dt = d p_i \wedge dq^i  - dH \wedge dt \ ,$$
showing   that $d \a^H$ is a source form, hence that   $\a^H$ is  of Poincar\'e-Cartan type. \par
 \smallskip
These observations show  that:
\begin{itemize}
\item[1)]{\it    Theorems  \ref{directNoether} and  \ref{inverseNoether} can be used for  $\cI_{[\a^H]}$
whenever   $\a^H$  is considered    on  a jet space $J^k(E)$ with  $k \geq 2$.}
\item[2)] {\it If   $\a^H$ is taken as a 1-form on $J^2(E)$ and we  consider adapted coordinates $(t, q^i_{(a)}, p^j_{(a)})_{a = 0,1,2}$,   the source form $\s$ in the variational class $[d \a^H]$ of the kind  \eqref{2.12}    is
\beq \label{sourceH} \s = \left(\frac{\partial H}{\partial q^i} + p_{(1)i}\right) \o^i_{(0)} \wedge dt + \left(\frac{\partial H}{\partial p_i} - q^i_{(1)}\right) \o_{(0)i} \wedge dt\ .\eeq
}
\item[3)] {\it The system given by the full prolongation of the Euler-Lagrange equations, determined by   \eqref{sourceH},  is
\beq\label{4.2} \left\{\begin{array}{ll}\s_i  = &\frac{\partial H}{\partial q^i} + p_{(1)i}  = 0\ ,\\
\ \\
 \wt \s^i  = & \frac{\partial H}{\partial p_i} - q^i_{(1)} = 0\ ,\ \\
 \ \\
 \frac{d \s_i}{dt} = & \frac{\partial^2 H}{\partial q^i q^j} q^j_{(1)} + \frac{\partial^2 H}{\partial q^i p_j} p_{(1)j }  + p_{(2)i} = 0\ ,\\
 \ \\
\frac{d \wt \s^i}{dt}  = &  \frac{\partial^2 H}{\partial p_i q^j} q^j_{(1)} + \frac{\partial^2 H}{\partial p_i p_j} p_{(1)j }  - q^i_{(2)} = 0\ .
\end{array}\right.
\eeq
}
\item[4)] {\it   $\a^H$ satisfies   condition (a)  of Theorem \ref{inverseNoether}}.
\item[5)] {\it $\a^H$   satisfies  also condition  (b)   of Theorem \ref{inverseNoether}, provided that it is  restricted to the open subset
$\cW = \{(H - p_i q^i_{(1)})(u_o) \neq 0\}$.}
\end{itemize}
\par
\bigskip
Due to this and Proposition \ref{lemma62}, given a  $(2n+1)$-tuple $\vv = (\vv^0, \vv^i, \vv_j)$ of smooth functions on a subset $\cW \subset J^2(E)$, the  $\cD$-symmetry
$$X_\vv \= \vv^0\frac{d}{d t} + \vv^i\frac{\partial}{\partial q^i}+ \vv_i\frac{\partial}{\partial p_i}+  \sum_{a=1}^2  \frac{d^a}{dt^a}\left(\vv^i_{(a)} - q^i_{(1)} \vv^0\right)\frac{\partial}{\partial q^i_{(a)}} +$$
$$ +   \sum_{a=1}^2  \frac{d^a}{dt^a}\left(\vv_{i} - p_{(1)i} \vv^0\right)\frac{\partial}{\partial p_{(a)i}}$$
is an infinitesimal symmetry for $\cI_{[\a^H]}$ if and only if  $\vv$ satisfies the  equation
\beq  p_{(1)i} \vv^i + p_i\frac{d \vv^i}{dt} - \frac{d H}{dt}\vv^0 - H \frac{d \vv^0}{dt} =  $$
$$ = -  \left(\frac{\partial H}{\partial q^i} + p_{(1)i}\right) (\vv^i  - q^i_{(1)} \vv^0) - \left(\frac{\partial H}{\partial p_i} - q^i_{(1)}\right) (\vv_i  - p_{(1)i} \vv^0)\ .\eeq

In addition, by Theorem \ref{inverseNoether},
given a constant of motion $f$ on $\cW \subset J^2(E)$, of order less than or equal to $1$,  we may locally  determine a $(2n+1)$-tuple $\vv$,  corresponding to
an infinitesimal symmetry $X_\vv$ for  $\cI_{[\a^H]}$ and such that
$$\imath_{X_\vv} \a^H =  f + g \ ,$$
where $g$ is a constant of motion that vanishes identically along the solutions of the variational principle. By the proof of Theorem \ref{inverseNoether}, the constant of motion $g$ (identically vanishing on solutions) and the infinitesimal symmetry $X_\vv$  are determined by the following steps: \\
\ \\
{\bf Step 1.} Find  smooth functions $(\wh \vv^i, \wh \vv_j, \wh \vv^i_{(1)}, \wh \vv_{j(1)}) $ such  that
$$- \frac{df}{dt} = \wh \vv^i \left(\frac{\partial H}{\partial q^i} + p_{(1)i}\right) + \wh \vv_i\left(\frac{\partial H}{\partial p_i} - q^i_{(1)}\right) +$$
$$ + \wh \vv^i_{(1)}\left(\!\frac{\partial^2 H}{\partial q^i q^j} q^j_{(1)} + \frac{\partial^2 H}{\partial q^i p_j} p_{(1)j }  + p_{(2)i}\! \right) + \wh \vv_{(1)i} \left(\! \frac{\partial^2 H}{\partial p_i q^j} q^j_{(1)} + \frac{\partial^2 H}{\partial p_i p_j} p_{(1)j }  - q^i_{(2)}\!\right)\!\!.$$
Note that {\it such functions do exist, but  are not uniquely determined by $f$}.
\\
\ \\
{\bf Step 2.} Determine the constant of motion $g$   by the formula
$$g \= \wh \vv^i_{(1)} \left(\frac{\partial H}{\partial q^i} + p_{(1)i}\right) + \wh \vv_{(1)i }\left(\frac{\partial H}{\partial p_i} - q^i_{(1)}\right)\ .
$$
Then the infinitesimal symmetry $X_\vv$ is
determined by   the $(2n+1)$-tuple $\vv$
\beq \left\{\begin{array}{ll}\vv^0  \= &\frac{1}{p_i q^i_{(1)} - H}\left( -   \sum_{i =1}^n\left(\wh \vv^i - \frac{d \wh \vv^i_{(1)}}{dt} \right)   p_i +  f + g \right)\ ,\\
\ & \ \\
\vv^i  \= & \wh \vv^i - \frac{d \wh \vv^i_{(1)}}{dt} + q^i_{(1)} \vv^0\ ,\\
\ & \ \\
 \vv_i \= & \wh \vv_j - \frac{d \wh \vv_{j(1)}}{dt} + p_{j(1)} \vv^0
 \end{array}
 \right.
 \eeq
and it is therefore equal to
 \beq X_\vv = \vv^0 \frac{d}{dt} + \left(\wh \vv^i - \frac{d \wh \vv^i_{(1)}}{dt} \right)\frac{\partial}{\partial q^i} + \left(\wh \vv_j - \frac{d \wh \vv_{j(1)}}{dt} \right) \frac{\partial}{\partial p_j} +\phantom{aaaaaaaaaaaaaa}$$
 $$ \phantom{aaaaa} +  \sum_{a=1}^2  \frac{d^a}{dt^a}\left(\wh \vv^i - \frac{d \wh \vv^i_{(1)}}{dt}\right)\frac{\partial}{\partial q^i_{(a)}} +  \sum_{a=1}^2  \frac{d^a}{dt^a}\left(\wh \vv_i - \frac{d \wh \vv_{i(1)}}{dt}\right)\frac{\partial}{\partial p_{i(a)}}\eeq

\bigskip

\subsection{Infinitesimal symmetries and $H$-symplectic symmetries}
\label{section43}
Consider now a constant of motion $f$ on $J^2(E)$ for the action $\cI_{[\a^H]}$ of order $0$ and time-independent, i.e. which is the pull-back  on $J^2(E)$ of
a function $\wt f: \cU \subset T^*M \longrightarrow \bR$, with $d \wt f\left(\frac{\partial }{\partial t} \right) = 0$.
We call any such constant of motion {\it first integral of elementary type}.
\par
\medskip
Now,  recall that  the system of equations that are satisfied by the solutions of the variational principle of $\cI_{[\a^H]}$ are \eqref{4.2}. It is then clear that,  for any $x_o \simeq (t_o, q^i_o, p_{oj}) \in T^*M$, there exists a jet
$$u_o \simeq (t_o, q^i_o, p_{oj}, q^i_{(a)o}, p_{oj(a)}) \in J^2(E)\ ,$$
satisfying  the full prolongation of Euler-Lagrange equations and  a solution $\g: I \subset \bR \longrightarrow E$, such that $\g_{t_o} = u_o$. The coordinates $q^i_{(1)o}, p_{j(1)o}$ of such point $u_o$ are equal to
$$q^i_{(1)o} = \left.\frac{\partial H}{\partial p_i}\right|_{(q^i_o, p_{oj}) }\ ,\qquad p_{j(1)o} = \left.\frac{\partial H}{\partial q^i}\right|_{(q^i_o, p_{oj}) }\ .$$

\par
Since the first integral $f$ depends only on the coordinates  of $T^*M$,  it follows that
\beq \label{4.5} 0 = df|_{u_o} \left(\dot \g^{(2)}_{t_o}\right)= \left.\frac{\partial f}{\partial q^i}\right|_{(q^i_o, p_{oj}) }\g^i_{(1)}(t_o) + \left.\frac{\partial f}{\partial p_i}\right|_{(q^i_o, p_{oj})}\g_{(1)j}(t_o) =  $$
$$ = \left.\frac{\partial f}{\partial q^i}\right|_{(q^i_o, p_{oj}) } \left.\frac{\partial H}{\partial p_i}\right|_{(q^i_o, p_{oj}) }- \left.\frac{\partial f}{\partial p_i}\right|_{(q^i_o, p_{oj}) }\left.\frac{\partial H}{\partial q^i}\right|_{(q^i_o, p_{oj}) }\eeq
for any $(q^i_o, p_{oi}) \in T^*M$. By arbitrariness of $(q_o^i, p_{oj})$, it follows that
$$- \frac{d f}{dt} = \frac{\partial f}{\partial q^i} q^i_{(1)} +  \frac{\partial f}{\partial p_j} p_{(1)j}  \overset{\eqref{4.5}}= \frac{\partial f}{\partial q^i} \left(q^i_{(1)} - \frac{\partial H}{\partial p_i}\right) +  \frac{\partial f}{\partial p_j} \left(p_{(1)j} + \frac{\partial H}{\partial q^i}\right) \ .$$
This shows that Step 1 of previous section can be easily solved by setting
$$\wh \vv^i = \frac{\partial f}{\partial p_i}\ ,\qquad \wh \vv_j = - \frac{\partial f}{\partial q^j} \ ,\qquad \vv^i_{(1)} = \vv_{(1)j} = 0\ .$$
From this, following Step 2,  we get that the (vanishing along solutions) constant of motion $g$ is identically  vanishing and that the infinitesimal symmetry $X^{(f)}$ associated with $f$ is
 \beq \label{uno} X^{(f)} =\frac{1}{p_i q^i_{(1)} - H}\left( -  \left( \frac{\partial f}{\partial p_i}\right)   p_i +  f  \right)\frac{d}{dt} +   \left(\frac{\partial f}{\partial p_i} \right)\frac{\partial}{\partial q^i} -  \left(\frac{\partial f}{\partial q^j} \right)  \frac{\partial}{\partial p_j} + $$
 $$ +  \sum_{a=1}^2  \frac{d^a }{dt^a}\left(\frac{\partial f}{\partial q^j} \right)\frac{\partial}{\partial q^i_{(a)}} -  \sum_{a=1}^2  \frac{d^a }{dt^a}\left(\frac{\partial f}{\partial q^j} \right)\frac{\partial}{\partial p_{i(a)}}\ .\eeq
\par
\bigskip
Consider now the natural immersion
$$\imath: E = T^*M \times \bR \longrightarrow J^2(E) \ ,\qquad \imath(\b, t) = j^2_t(\wt \b)\ ,$$
where $\wt \b: \bR \to E$ denotes the constant section $\wt \b(t) \equiv \b$,
and the projection
$$ \wh \pi: E = T^*M \times \bR \longrightarrow T^*M\ .$$
For any infinitesimal symmetry \eqref{uno}, consider the associated vector field $Y^{(f)} $ on (an open subset of) $T^*M$,  defined by
\beq\label{due}Y^{(f)} \= \wh \pi_*\left(\left.X^{(f)}\right|_{\imath(E)}\right) =   \left(\frac{\partial f}{\partial p_i} \right)\frac{\partial}{\partial q^i} -  \left(\frac{\partial f}{\partial q^j} \right)  \frac{\partial}{\partial p_j}\ . \eeq
Notice that:
\begin{itemize}
\item[i)] The vector field $Y^{(f)}$ is the {\it Hamiltonian vector field associated with $f$}, i.e. the unique vector field on $T^*M$ that satisfies the equation
\beq \imath_{Y^{(f)}} \O = df\ .\eeq
Note also that, by  \eqref{4.5}, it  satisfies  the condition
\beq \label{lilletta} \imath_{Y^{(f)}} dH = 0\ .\eeq
\item[ii)] Conversely,  if $Y^{(f)}$ is a Hamiltonian vector field on $\cU \subset T^*M$, associated with a function $f$ on $\cU$ and satisfying \eqref{lilletta}, one can  directly check that $f: \cU \longrightarrow \bR$ is a first integral of elementary type.
\item[iii)] In any open subset $\cW \subset J^2(E)$, where a system of adapted coordinates $\wh \xi^{(2)} = (t, q^i, p_j, q^i_{(1)}, p_{j(1)}, q^i_{(2)}, p_{(2)j)})$ are defined, given a Hamiltonian vector field as in (ii), there is a unique infinitesimal symmetry $X^{(f)}$ for the variational principle of  $\cI_{[\a^H]}$ of the form \eqref{uno} and such that $Y^{(f)} \=  \wh \pi_*\left(\left.X^{(f)}\right|_{\imath(E)}\right)$.
\item[iv)] Given $u = \imath(\b_o) \in \imath(E) \subset J^2(E)$, the correspondence
$X \longmapsto \wh \pi_*\left(\left.X\right|_{\imath(E)}\right)$  determines an isomorphism between the Lie algebra of germs at   $\b_o$ of the infinitesimal symmetries  as in  \eqref{uno} and the Lie algebra  of germs at  $\b_o$ of the  Hamiltonian vector fields on $T^*M$, which satisfy \eqref{lilletta}.
\end{itemize}
\medskip
These facts can be nicely summarized using the following notion.
\begin{definition} The vector fields $Y$ on $\cU \subset T^*M$ that satisfy the equations
\beq\cL_Y \O = 0\ , \qquad   \imath_Y d H = 0\eeq
are called {\it  $H$-symplectic symmetries of $T^*M$}
\end{definition}
\par
\bigskip
Fix now a point $u \in \cU \subset T^*M$ and consider the following classes of germs at $u$:
$$\begin{array}{l}
\sp^{H} \=  \{\ \text{germs at}\ u \ \text{of $H$-symplectic symmetries}\ \}\ ,\\
\ \\
\FirstEl\= \{\ \text{germs at}\ u \ \text{of first integrals  for}\ \cI_{[\a^H]}\ \text{of elementary  type}\ \} \ .\\
\end{array}
$$
By the above discussion, the  correspondence between infinitesimal symmetries and constants of motion, given by Noether Theorems,  determine the
 isomorphism of vector spaces
\beq \varphi: \sp^{H} \longrightarrow \FirstEl/\bR\ , \eeq
where, for any $\underline Y \in \sp^{H}$, the corresponding equivalence class $\varphi(\underline Y) \in \FirstEl/\bR$ is determined by the  $\underline f$ (determined up to an additive  constant) such that
$$\left.\imath_Y \O\right|_{u}  = \left.d f\right|_{u}\ .$$
Since $\sp^{H}$ has a natural structure of Lie algebra, the vector space isomorphism $\varphi$ induces a natural  Lie algebra structure
on $\FirstEl/\bR$. \par
\medskip
We remark   that {\it the Lie brackets of the induced Lie algebra structure  are the usual {\rm Poisson brackets} of the symplectic manifold  $(T^*M, \O)$}.
\par
\bigskip
\subsection{The infinite-dimensional Lie algebra $\sp^{H}$}
 Let  $\O_o$ be the standard symplectic form of $\bR^{2n}$, i.e.
\beq \O_o = dx^1 \wedge dx^2 + dx^3 \wedge dx^4 + \ldots + dx^{2n-1} \wedge dx^n\ ,\eeq
 and denote by $\sp_{\infty}^{(1)}(2n,\bR)$ the Lie subalgebra of $\sp_{\infty}(2n,\bR)$, determined by  the   vector fields,
 commuting with $\frac{\partial}{\partial x^1}$. We recall that $\sp_{\infty}(2n,\bR)$ is the infinite-dimensional Lie algebra of the germs at $0$ of vector fields of  $\bR^{2n}$, which preserve $\O_o$. Consequently,  the (infinite-dimensional) Lie algebra   $\sp_{\infty}^{(1)}(2n,\bR)$  is made  of  all germs of  vector fields $X$
satisfying the pair of conditions
\beq \label{4.14}\cL_X \O_o = 0\qquad \text{and}\qquad\cL_{\frac{\partial}{\partial x^1}} X = 0\ .\eeq
 Note that  the second condition in  \eqref{4.14} is   equivalent to require that
$$dx^2(X) = \cL_X d x^2 =  \cL_X (\imath_{\frac{\partial}{\partial x^1}} \O_o)= 0\ .$$
One can directly check that $\underline X \in \sp_{\infty}^{(1)}(2n,\bR)$ if and only if $X$  is of the form
$$X = h \frac{\partial}{\partial x^1} + \wt X\qquad \text{with}\qquad \wt X = \sum_{j = 3}^{2n} \wt X^j \frac{\partial}{\partial x^j}$$
where $h$ and $\wt X^i$ are   functions that satisfy the equations
$$\frac{\partial h}{\partial x^1} = \frac{\partial \wt X^j}{\partial x^1} = 0\ ,\qquad \cL_{\wt X}  \O_o'  = - dh \wedge dx^2\ ,$$
where $ \O'_o$ denotes the standard symplectic form of $\bR^{2n-2} = \{\ x \in \bR^{2n}\ :\ x^1 = x^2 = 0\ \}$.\par
\bigskip
 This means that
$$\sp^{(1)}_\infty(2n, \bR) \supset \sp_\infty(2n-2, \bR)\times \bR\ . $$
\par
\bigskip

Let $H: \cU \subset T^*M\longrightarrow \bR$ be a time-independent Hamiltonian. An element  $\b \in \cU$ is called {\it point of non-degeneracy for $H$} if $dH|_\b \neq 0$.
\begin{prop} For any Hamiltonian $H$ and  any point of non-degeneracy $u$ for $H$, the Lie algebra $\sp^{H_o} $ is isomorphic to the infinite dimensional Lie algebra $\sp^{(1)}_\infty(2n, \bR)$.
\end{prop}
\begin{pf} By the proof of Darboux Theorem (see e.g. \cite{Ar}), since $dH|_u \neq 0$, there exists a system of coordinates around $u$, in which $ \O$ assumes the same expression of the standard symplectic form $ \O_o$ and the function $H$ is equal to $H = x^2$. From this, the conclusion follows.
\end{pf}
\par
From the above proposition,  around  points of non-degeneracy, all Lie algebras of (germs of) first integrals of elementary type of all Hamiltonians are  infinite-dimensional and mutually isomorphic.  The same clearly occurs for any  subalgebra $\gg$ of such Lie  algebras and gives rise   to the following  phenomenon (see also \cite{Mu} for a  constructive proof of this property for some special Lie algebras).
\par
\begin{theo} \label{theorem44}  {\it Assume that   for a given  Hamiltonian $H$ there exists  a collection of first integrals of elementary type, which (by means of Poisson brackets) constitutes  a (finite or infinite) dimensional Lie algebra  $\gg \subset \sp^{H} $  at a point of non-degeneracy $u$.\par
Then  the same occurs       for any other Hamiltonian $H'$ in the following sense:  around any  point of non-degeneracy of $H'$,  there exists    a collection of (locally defined!)  first integrals of elementary type for $H'$, which  constitutes  a  Lie algebra  $\gg'  \subset \sp^{H'} $ that  is isomorphic to $\gg$.}
\end{theo}



\begin{thebibliography}{11}

\bibitem[1]{Ar} V.I. Arnold, Mathematical Methods of Classical Mechanics (2nd ed.), {\it Springer-Verlag, New York}, 1989.

\bibitem[2]{Ge} S. Germani, {\it Leggi di conservazione e simmetrie: un approccio geometrico al Teorema di Noether},
Tesi di Laurea Magistrale, {\it Universit\`a di Camerino, Camerino}, 2012.

\bibitem[3]{FGS} E. Fiorani, S. Germani  and A. Spiro, {\it On  the Lie algebras
of conservation laws of  variational  partial differential equations}, in preparation.

\bibitem[4]{Ko0} Y. Kosmann-Schwarzbach, {\it On the momentum mapping in field theory}, in ``Differential geometric methods in mathematical physics (Clausthal, 1983)'', p. 25--73, Lecture Notes in Math., {\bf 1139}, {\it Springer, Berlin}, 1985.

\bibitem[5]{Ko} Y. Kosmann-Schwarzbach, The Noether Theorems, {\it Springer, New York}, 2011.

\bibitem[6]{MR} A. E. Marsden and T. S. Ratiu, Introduction to Mechanics and Symmetry, {\it Springer-Verlag, Berlin}, 1994.

\bibitem[7]{Mi} J. Milnor, Morse Theory, {\it Princeton Univ. Press, Princeton}, 1963.

\bibitem[8]{Mu} N. Mukunda, {\it Dynamical Symmetries and Classicam Mechanics}, Phys. Rev., {\bf 155} (1967), 1383--1386.

\bibitem[9]{No} E. Noether, {\it Invariante Variationsprobleme}, Nachr. K\"{o}nig. Gesell. Wissen. G\"{o}ttingen, Math.-Phys.
Klasse, (1918), 235--257.

\bibitem[10]{Ol} P. J. Olver, Applications of Lie Groups to Differential Equations, {\it Springer-Verlag, New York}, 1986; 2nd ed. revised, 1993.

\bibitem[11]{Ol1} P. J. Olver, {\it Noether's theorems and systems of Cauchy-Kovalevskaya type}, in ``Nonlinear systems of partial differential equations in applied mathematics, Part 2 (Santa Fe, N.M., 1984)'',
Lectures in Appl. Math., 23, {\it Amer. Math. Soc., Providence, RI, 1986.}

\bibitem[12]{Ol2} P. J. Olver, {\it The Noether theorems. Invariance and conservation laws in the twentieth century [book review of MR2761345]}, Bull. Amer. Math. Soc. (N.S.),  {\bf 50} (2013), 161--167.

\bibitem[13]{Sp} A. Spiro, {\it Cohomology ol Lagrange complexes invariant under pseudogroups ol local transformations},
Int. J. Geom. Methods Mod. Phys., {\bf 4} (2007), 669--705.

\bibitem[14]{Ya} K. Yamaguchi, {\it Contact geometry of higher order},  Japan. J. Math. (N.S.), {\bf 8} (1982),  109--176.

\bibitem[15]{Ya1} K. Yamaguchi, {\it Geometrization of jet bundles},  Hokkaido Math. J., {\bf 12} (1983), 27--40.

\end{thebibliography}
\end{document}